\documentclass{ieeeaccess}
\usepackage{algorithm} 
\usepackage{algpseudocode}
\usepackage{amsmath,amsfonts,amssymb}
\usepackage{amsthm}
\usepackage{caption}
\usepackage{cite}
\usepackage{graphicx}
\usepackage{textcomp}
\usepackage{subfigure}
\usepackage{multirow}
\usepackage{rotating}
\usepackage{adjustbox}
\usepackage{diagbox}
\usepackage{stfloats}
\usepackage{url}
\usepackage{bm}
\usepackage{longtable}
\usepackage{booktabs}
\usepackage{lscape}

\renewcommand{\figurename}{FIGURE}
\def\BibTeX{{\rm B\kern-.05em{\sc i\kern-.025em b}\kern-.08em
    T\kern-.1667em\lower.7ex\hbox{E}\kern-.125emX}}

\begin{document}
	
\history{Date of publication xxxx 00, 0000, date of current version xxxx 00, 0000.}
\doi{10.1109/ACCESS.2017.DOI}

\title{A Comprehensive Review for Breast Histopathology Image Analysis Using Classical and Deep Neural Networks}
\author{{Xiaomin Zhou\authorrefmark{1}}, 
{Chen Li\authorrefmark{1}}, 
{Md Mamunur Rahaman\authorrefmark{1}},
{and Yudong Yao (IEEE Fellow)\authorrefmark{2}},
{Shiliang Ai\authorrefmark{1}}, 
{Changhao Sun\authorrefmark{1}},
{Qian Wang\authorrefmark{3}},
{Yong Zhang\authorrefmark{3}},
{Mo Li\authorrefmark{3}},
{Xiaoyan Li\authorrefmark{3}},
{Tao Jiang\authorrefmark{4}},
{Dan Xue\authorrefmark{1}},
{Shouliang Qi\authorrefmark{1}},
{Yueyang Teng\authorrefmark{1}}
}
\address[1]{Microscopic Image and Medical Image Analysis Group, MBIE College, Northeastern University, 110169, Shenyang, China}
\address[2]{Department of Electrical and Computer Engineering, Stevens Institute of Technology, Hoboken, NJ 07030, USA}
\address[3]{Cancer Hospital of China Medical University, Liaoning Hospital and Institute, Shenyang 110042, China}
\address[4]{Control Engineering College,Chengdu University of Information Technology,Chengdu 610103,China}

\tfootnote{We acknowledge financial support from the ``National Natural Science Foundation of China'' (No. 61806047), the ``Fundamental Research Funds for the Central Universities'' (N2019003), the ``Scientific Research Fund of SiChuan Provincial Science \& Technology Department'' (No. 2017TD0019), the ``Scientific Research Fund of Chengdu Science and Technology Burea'' (No. 2017-GH02-00049-HZ, 2018-YF05-00981-GX) and the ``China Scholarship Council'' (No. 2018GBJ001757).}


\corresp{Corresponding author: Chen Li (e-mail: lichen201096@hotmail.com).}

\begin{abstract}
Breast cancer is one of the most common and deadliest cancers among women. Since histopathological images contain sufficient phenotypic information, they play an indispensable role in the diagnosis and treatment of breast cancers. To improve the accuracy and objectivity of \textit{Breast Histopathological Image Analysis} (BHIA), \textit{Artificial Neural Network} (ANN) approaches are widely used in the segmentation and classification tasks of breast histopathological images. In this review, we present a comprehensive overview of the BHIA techniques based on ANNs. First of all, we categorize the BHIA systems into classical and deep neural networks for in-depth investigation. Then, the relevant studies based on BHIA systems are presented. After that, we analyze the existing models to discover the most suitable algorithms. Finally, publicly accessible datasets, along with their download links, are provided for the convenience of future researchers.
\end{abstract}


\titlepgskip=-15pt
\maketitle
\section{Introduction}
\label{sec:introduction}
\PARstart{B}{reast} cancer is the most commonly diagnosed and leading cause of cancer deaths among women~\cite{bray-2018-GCSG}. According to the World Health Organization (WHO), every year 2.1 million women have breast cancer worldwide. In 2018, an estimated 627,000 women died, representing about $15\%$ of all cancer deaths among women\cite{WHO-2019-Date}. In the United States, it ranks first in the record of the most common cancers that women are expected to be diagnosed in 2019 at a rate of up to $30\%$~\cite{CS-2019-Date}.

\begin{figure}[htbp!]
  \centering
  \subfigure[]{
    \label{fig:subfig:two-stage-seg01} 
    \includegraphics[width=0.44\linewidth]{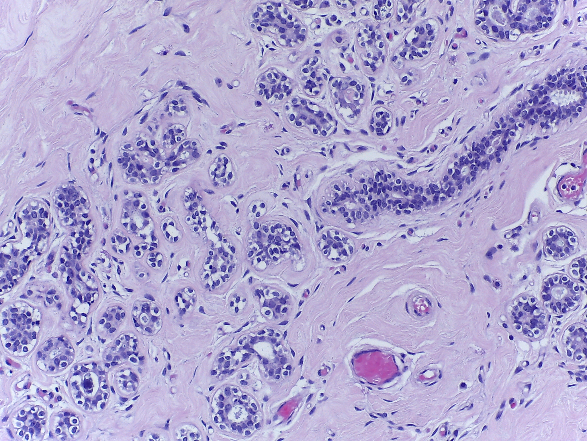}}
  \subfigure[]{
    \label{fig:subfig:two-stage-seg02} 
    \includegraphics[width=0.44\linewidth]{Benign.png}}
     \subfigure[]{
    \label{fig:subfig:two-stage-seg03} 
    \includegraphics[width=0.44\linewidth]{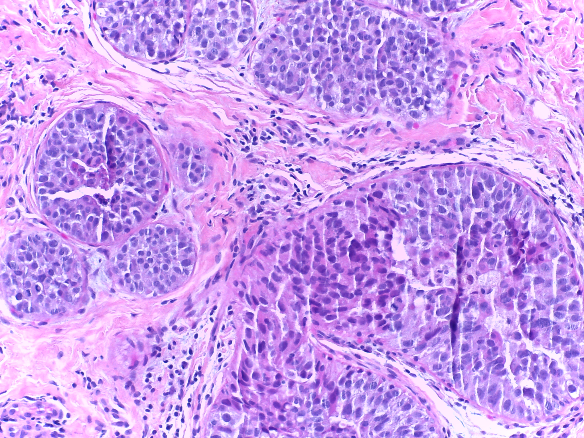}}
     \subfigure[]{
    \label{fig:subfig:two-stage-seg04} 
    \includegraphics[width=0.44\linewidth]{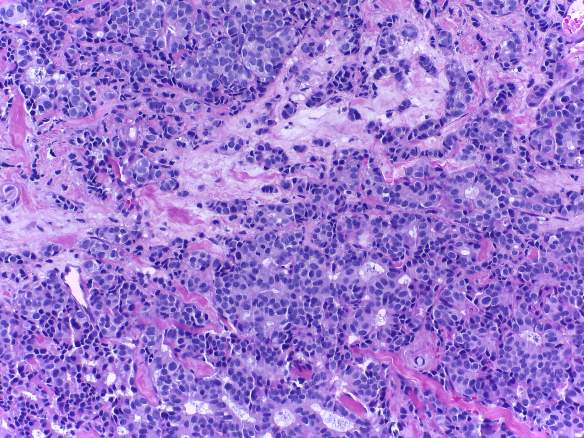}}
\caption{H\&E stained images of different type, (a) is normal tissue, (b) is benign abnormality, (c) is \textit{in-situ} carcinoma, and (d) is invasive carcinoma. These images are from the BACH dataset~\cite{aresta-2019-BGCB}.}
  \label{fig:example} 
\end{figure}

There are four types of breast tissue i.e., normal, benign, \textit{in-situ} carcinoma, and invasive carcinoma. Benign tissue refers to a minor change in the structure of the breast, but it is not classified as cancer, and in most cases, it is not harmful to health. \textit{In-situ} carcinoma remains in the mammary duct lobule system and does not affect other organs. If it diagnoses in time, \textit{in-situ} carcinoma can be cured. However, invasive carcinoma is a malignant tumor that tends to spread in other organs. There are many techniques for breast cancer detection, such as X-ray mammography~\cite{moghbel-2019-ARBB}, 3-D Ultrasound (US)~\cite{kozegar-2019-CADA}, Computed Tomography (CT), Positron Emission Tomography (PET)~\cite{domingues-2019-UDLT}, Magnetic Resonance Imaging (MRI)~\cite{murtaza-2019-DLBB}, and breast temperature measurement~\cite{moghbel-2013-ARCA}. However, pathological diagnosis is often regarded as a "golden standard"~\cite{Jonathan-2019-HIPA}. For better observation and analysis, the removed tissues usually need to be stained, where the \textit{Hematoxylin and Eosin} (H\&E) staining approach is the most common method. The hematoxylin dyes the nuclei a dark purple color and the eosin dyes other structures (cytoplasm, stroma, etc.) a pink color. Make it like~\figurename~\ref{fig:example}, showing the different types of breast tissue images stained with H\&E.

In histopathological research, the sections are examined under a microscope to analyze the characteristics and properties of the tissues by a histopathologist~\cite{Romos-Vara-2011-PAM}. Traditionally, the tissue sections are observed by the naked eyes of the histopathologist directly, and the visual information is analyzed based on the prior medical knowledge manually. However, due to the complexity and diversity of histopathological images, this manual analysis can take much time. At the same time, the objectivity of this manual analyzing process is unstable, depending on the experience, workload, and mood of the histopathologist greatly.

In recent years, \textit{Artificial Intelligence} (AI) technology develops rapidly. In particular, it makes important achievements in computer vision, image processing and analysis. AI also shows potential advantages in histopathological analysis. AI-assisted diagnosis can undertake tedious focus screening work and quickly extract valuable information related to diagnosis from massive data. Meanwhile, AI has a strong objective analysis ability in histopathological detection and can avoid subjective differences caused by manual analysis. To some extent, the misjudgment of pathologists can be reduced and the working efficiency can be improved.

 \subsection{General development of existing AI analysis Histopathology }
\label{ss:Introduction:General}
\begin{figure*}[htbp!]
\centering
 \includegraphics[width=1\linewidth]{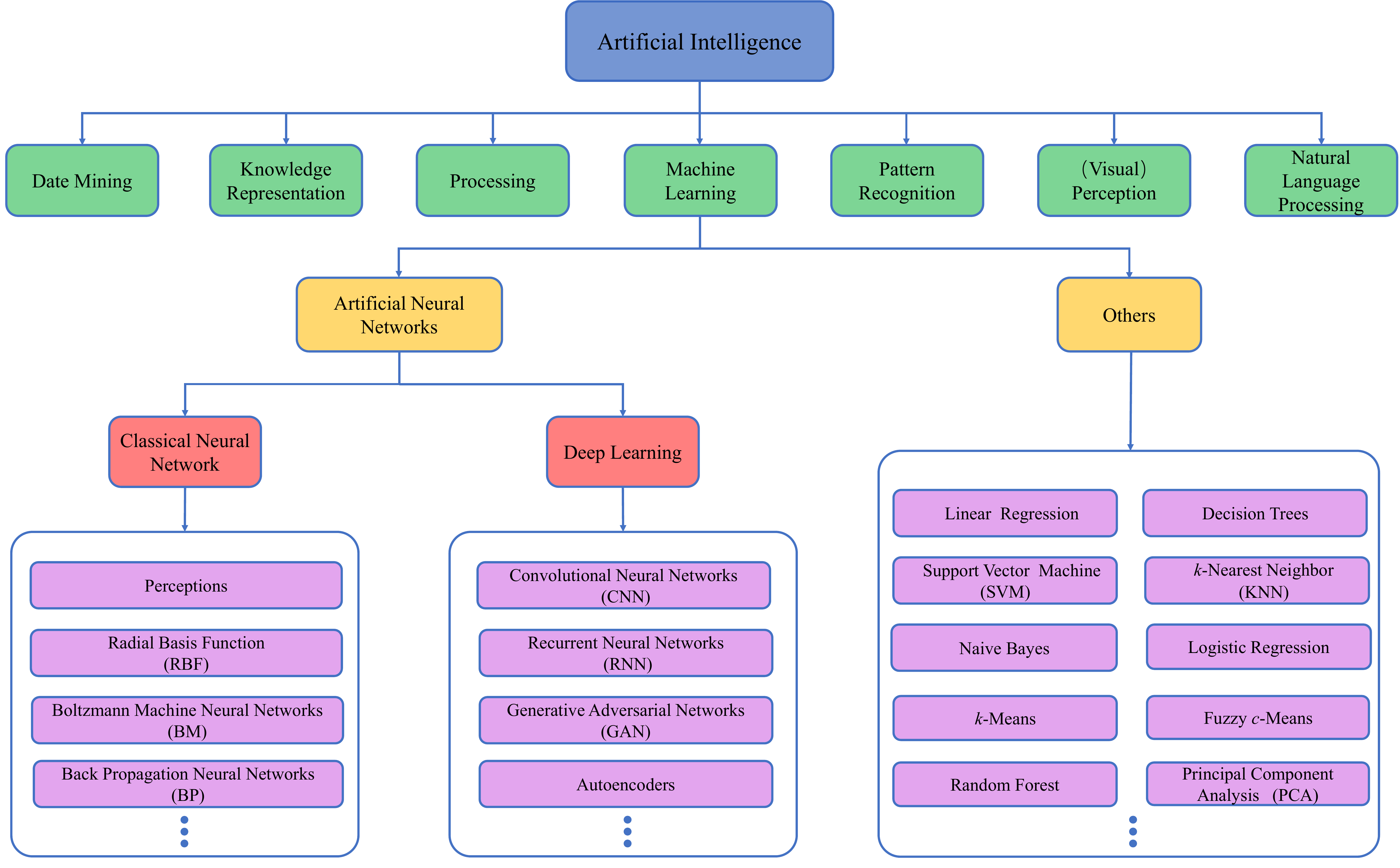}
\caption{The structure of ANN technology in the AI knowledge system.}
\label{fig:Structure}
\end{figure*}

AI is an umbrella term encompassing the techniques for a machine to mimic or go beyond human intelligence, mainly in cognitive capabilities~\cite{robertson-2018-DIAB}. As shown in \figurename~\ref{fig:Structure}, the main contents of AI research include \textit{Machine Learning} (ML), pattern recognition, natural language processing, etc. Especially, ML has a significant contribution to the development of medicine.

ML is used in the pathological diagnosis of different cancer fields, e.g., cervical cancer, gastric cancer, colon cancer, lung cancer and breast cancer. The scope of the application focuses on the benign and malignant diagnosis, disease grading, staining analysis, and early tumor screening. For example, the work of~\cite{Lichen-2019-CHIC} proposes a weakly supervised multi-layer hidden conditional random field model to classify the cervical histopathological images into well, moderate and poorly differentiated stages. In the experiment, the proposed method is tested on the six cervical IHC datasets and obtains an overall classification accuracy of $77.32\%$ and the highest one of the six is $88\%$, showing the effectiveness and potential of the method. In the field of gastric cancer, the work of~\cite{liYuexiang-2018-DLBG} proposes a deep learning based framework, namely GastricNet, for automatic gastric cancer identification. The experimental results show that this deep learning framework performs better than state-of-the-art networks like DenseNet, ResNet, and achieves an accuracy of $100\%$ for slice-based classification.  The colorectal cancer is a malignant tumor that starts in the form of growth known as polyps mainly in the inner linings of the colon or rectum part.  In the work of~\cite{Dabass-2018-FGCC}, proposing an automated supervised technique using deep learning to keep original image size is proposed in this paper for doing five-grade cancer classification via 31 layers \textit{Deep Convolutional Neural Network} (DCNN). The proposed model results in classification accuracy of $96.97\%$ for two-class grading and $93.24\%$ for five-class cancer grading. About lung cancer pathology, Adenocarcinoma (LUAD) and squamous cell carcinoma (LUSC) are the most prevalent subtypes of lung cancer. The study of~\cite{coudray-2018-CMPN} trains a DCNN (inception-v3) on \textit{Whole Slide Images} (WSIs) obtained from The Cancer Genome Atlas to accurately and automatically classify them into LUAD, LUSC or normal lung tissue, with an average \textit{Area Under the Curve} (AUC) of 0.9.

\begin{figure}[htbp!]
\centering
 \includegraphics[width=1\linewidth]{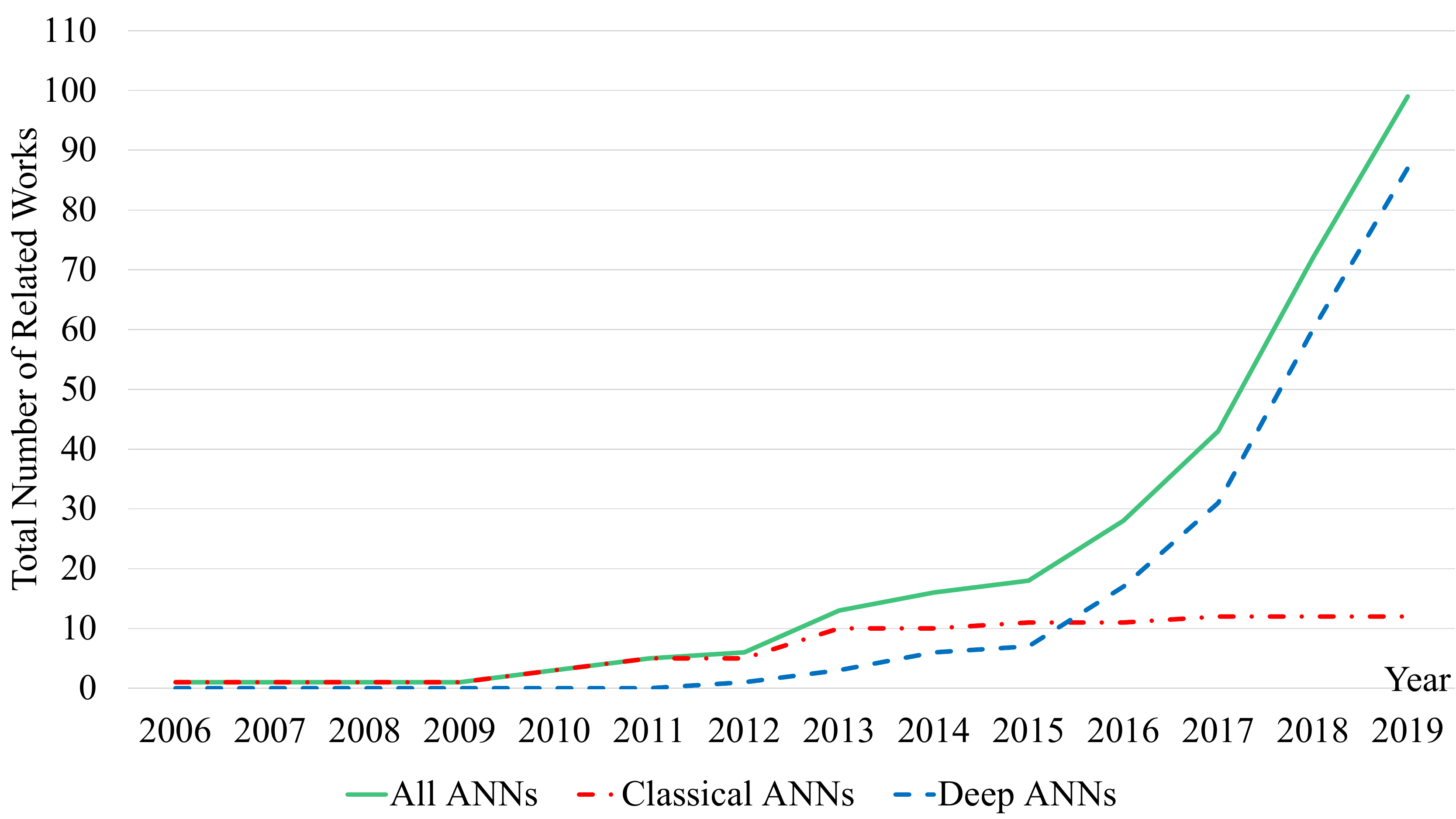}
\caption{The development trend of ANN methods for BHIA tasks. 
The horizontal direction shows the time. The vertical direction shows the cumulative number of related works in each year.}
\label{fig:Trend2019}
\end{figure}

It is worth noting that ANNs, as a branch of machine learning, play an important role in pathological diagnosis. ANN method, including classical and deep neural networks, is a kind of mathematical model or calculation model which imitates the structure and function of the biological neural network. In recent years, ANNs are widely used in \textit{Breast Histopathological Image Analysis} (BHIA) for image segmentation, feature extraction, and classification. The development trend of BHIA using ANNs is shown in~\figurename\ref{fig:Trend2019}.

\subsection{Motivation of our review paper }
\label{ss:Introduction:Motivation}
This paper focuses on the work of ANNs in the image analysis of breast histopathology. A comprehensive overview of techniques for image analysis of breast histopathology using classical neural networks and deep neural networks is presented. The motivation is to clarify the development history of ANNs, understand the popular technology and trend of ANN applications, and discover the future potential of ANNs in the BHIA field. As far as we know, there exist some survey papers that summarize papers related to the BHIA work (e.g., the reviews in~\cite{Jonathan-2019-HIPA,robertson-2018-DIAB,Gil-2002-IAA,Demir-2005-ACD,Gurcan-2009-HIA,He-2010-CAD,He-2012-HIA,Irshad-2014-MFN,Veta-2014-BCH,Bhattacharjee-2014-ROH,Arevalo-2014-HIR,Aswathy-2017-DOB,Chen-2017-CPO,Steiner-2018-IOD,Acs-2018-NJD,hamidinekoo-2018-DLMB,debelee-2019-SDLB,kaushal-2019-RTCA,murtaza-2019-DLBC,li-2019-ASBH}). In the following part, we go through the survey papers that are related to the BHIA work.

The survey of~\cite{Jonathan-2019-HIPA} reviews machine learning methods that are usually employed in histopathological image processing, such as segmentation, feature extraction, unsupervised learning, and supervised learning. More than 130 papers about histopathological image analysis are summarized, but only five are about BHIA with ANNs.

The survey of~\cite{robertson-2018-DIAB} publishes a research survey, focusing on the use of AI and deep learning in the diagnosis of breast pathology images, and other recent developments in digital image analysis. Among them, we are only interested to summarize the development of deep learning in breast pathological diagnosis from the application direction. However, the article does not discuss the results that have been obtained for each research method.

The survey of~\cite{hamidinekoo-2018-DLMB} summarizes current deep learning techniques in mammography and breast histology. In this article we only focus on deep learning techniques on breast histopathology images. According to different tasks, namely nuclei analysis, tubular analysis, epithelial and stromal region analysis, mitotic activity analysis and other tasks in the breast digital histopathology image process, 16 papers are summarized based on BHIA.

The survey of~\cite{debelee-2019-SDLB} summarizes the deep learning applications in breast cancer image analysis in Screen-File Mammography (SFM), Digital Mammography (DM), US, MRI, and Digital fast tomosynthesis (DBT) imaging modes, respectively. At the same time, six papers are found based on the topic of our interest.

In~\cite{kaushal-2019-RTCA}, an overview of ``recent trends in computer assisted diagnosis system for breast cancer diagnosis using histopathological images'' with 106 related works is presented. This review summarizes those works by four technical steps, including image pre-processing, segmentation, feature extraction and selection, as well as classification. However, there are only 20 related works about BHIA with ANNs in this paper.

The survey of~\cite{murtaza-2019-DLBC} publishes a review about the classification tasks of breast cancer on deep learning. The author reviews five aspects, namely datasets used, various medical imaging modalities exploited, image pre-processing techniques, types of DNNs, and the performance metrics used to construct and evaluate breast cancer classification models. The paper cites 49 studies, of which 27 are about histopathological images, and the rest are about mammograms. In the empirical evaluation, the analysis based on histopathological images and mammograms is not distinguished, and the time is only 2014-2018.

In our previous work~\cite{li-2019-ASBH}, we propose a  brief survey for breast histopathology image analysis using classical and deep neural networks. With more than 60 related works, referring to classical ANNs, deep ANNs and methodology analysis.

Besides our previous brief review about BHIA with ANN techniques, there is not a special one that focuses on the ANN approaches in this field. Hence, in order to clarify the BHIA work using ANN approaches in recent years, as of early 2020, based on our previous work in~\cite{li-2019-ASBH}, we summarize more than 150 related works to prepare this comprehensive review. We propose this paper with the following structure: In Sec.~\ref{s:classical}, the BHIA work using classical ANN methods are introduced; in Sec.~\ref{s:deep}, the state-of-the-art deep ANN methods are summarized; Sec.~\ref{s:method}, presents the method analysis; Sec.~\ref{s:conc} concludes this paper and discusses the future work.

\section{BHIA Using Classical ANNs}
\label{s:classical} 
An overview of the BHIA work using classical ANN methods is compiled in this section. Then, we analyze and summarize the chapter.
\subsection{Related Works}
\label{ss:classical:related}
In this section, we divide related work into classification tasks and segmentation tasks according to the motivation. Then, we summarize the contributions, methods, and results of each paper. 
\paragraph{\textbf{Classification Tasks: }}

In~\cite{Petushi-2006-LCO}, in order to evaluate two proposed texture features, a third-party software (LNKnet package) containing a neural network classifier is used. In the experiment, 536 samples are used for classifier training and 526 samples are used for testing. Finally, an accuracy of $90\%$ is achieved.

In~\cite{osareh-2010-MLTD}, \textit{Support Vector Machine} (SVM), \textit{$k$-Nearest Neighbor} (KNN) and \textit{Probabilistic Neural Networks} (PNN) classifiers are combined with signal-to-noise ratio feature ranking, sequential forward selection-based feature selection and principal component analysis feature extraction to distinguish the benign and malignant tumors of the breast. Finally, the best overall accuracies for breast cancer diagnosis are achieved by using an SVM classifier. The accuracy of  $98.80\%$ is achieved on dataset 1 (692 specimens of fine-needle aspirates of breast lumps), and $96.33\%$ is achieved on dataset 2 (295 microarrays). Similarly, PNN achieves $97.23\%$ and $93.39\%$ overall accuracy on dataset 1 and dataset 2, respectively.

In~\cite{Singh-2011-BCD}, four types of H\&E stained breast histopathology images are classified, using eight features and a three-layer forward/back ANN classifier. 
In the experiment, 1808 training samples, 387 validation samples, and 387 test samples are tested, and an overall accuracy around $95\%$ is achieved.

In~\cite{Zhang-2011-BCC,Zhang-2013-BCD,Zhang-2013-BCH}, an automatic breast cancer classification scheme based on histopathological images is proposed. First, edge, texture and intensity features are extracted. Then, based on each of the extracted features, an ANN classifier is designed, respectively. Thirdly, an ensemble learning approach, namely ``random subspace ensemble'', is used to select and aggregate these classifiers for even better classification performance. Finally, a classification accuracy of $95.22\%$ is obtained on a public image dataset.

In~\cite{Loukas-2013-BCC}, in order to classify low magnification ($10\times$) breast cancer histopathology images (H\&E stained) into three malignancy grades, 30 texture features are extracted first. Then, feature selection is applied to find more effective information from the extracted features. Thirdly, a PNN classifier is built up based on the selected features. Lastly, 65 images are tested in the experiment, and an overall accuracy around $87\%$ is obtained.

In~\cite{shukla-2017-CHIB}, morphological features are extracted to realize the classification of cancerous and non-cancerous cells in histopathological images, and 70 histopathological images are randomly selected as the dataset. In the experiment, a multi-layer perceptron, based on feed-forward artificial neural network modal, achieves $80\%$ accuracy, $82.9\%$ sensitivity and $89.2\%$ AUC, respectively. 

\paragraph{\textbf{Segmentation Tasks: }} 
In~\cite{Kowal-2013-CDO}, a competitive neural network is applied as a clustering based method to segment breast cancer regions from needle biopsy microscopic images. In this work, 21 shape, texture and topological features are extracted. Then, the network is used to cluster the images into different regions based on these features. In the experiment, a dataset with over 500 images is tested, and an overall accuracy of around $98.7\%$ is achieved.

\begin{figure}[htbp!]
\centering
\includegraphics[width=1\linewidth, height=4.5cm]{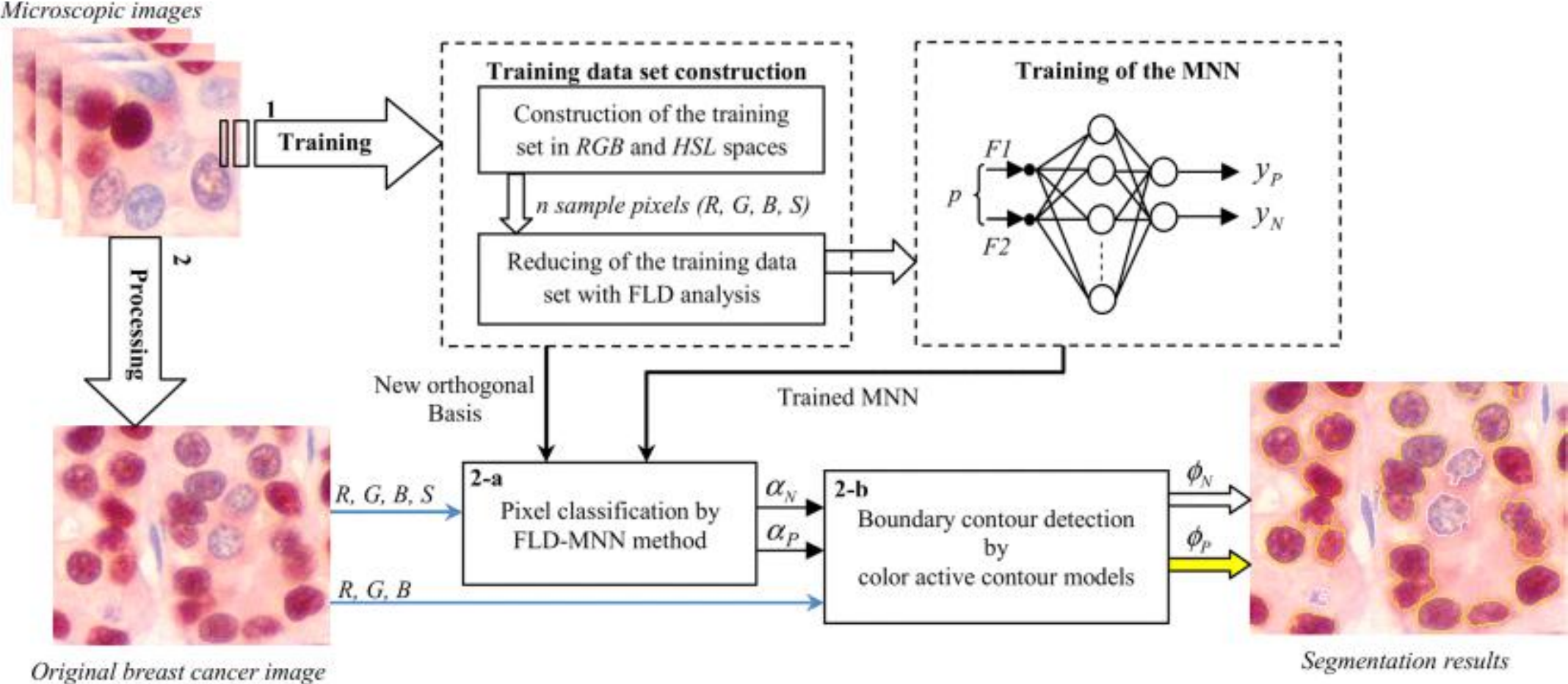}
\caption{Flow chart of the proposed segmentation scheme for cancer nuclei detection in~\cite{Mouelhi-2013-ASS}. The yellow contour represents the outline of the identified positive nucleus, while the white contour represents the outline of the identified negative nucleus. This figure corresponds to Fig.2 in the original paper.}
\label{fig:seg}
\end{figure}

In~\cite{Mouelhi-2013-ASS}, a supervised segmentation scheme using multilayer neural network and color active contour model to detect breast cancer nuclei is proposed. In this work, 24 images are used to test the method, and an average accuracy of $95.5\%$ is finally achieved. The flow chart is shown in \figurename~\ref{fig:seg}.

\subsection{Summary}
\label{ss:classical:summary}

According to the review above, we can see that the ANNs used in the BHIA field around 2012 are classical neural networks. The classical neural network has remarkable performance in various fields, but it also has some limitations, such as easier to overfit, slow training speed, and can only set parameters according to experience. Due to the low computational speed of the computer and the lack of sufficient data to train the computer system at the time, it is impossible to extract the effective ANN features from the raw data. Therefore, most of the classical neural networks in the field of BHIA are used as classifiers. In the aspect of feature selection, most research works use texture features and morphological features for segmentation and classification. \tablename~\ref{table1} summarizes the work of different teams using classical neural networks in analyzing histopathological images of breast cancer. Further details of the method analysis are discussed in Sec.~\ref{ss:methods_classical}

\begin{table*}[htbp!]
\renewcommand\arraystretch{1.25}
\setlength{\tabcolsep}{1 pt}
\centering
\scriptsize\caption{Histopathology and classical ANNs based breast cancer image analysis. Multi-Layer Perceptron (MLP), Probabilistic Neural Networks (PNN), Multi-layer Neural Network (MNN), Accurcy (Acc), and Sensitivity (Sn). The second column ``Detail'', shows the number of classes and segmentation regions.}
\begin{tabular}{|c|c|c|c|c|c|c|c|}
\hline 
Aim                              & Detial                  & Year                  & Reference                           & Team                                  & Data Information                                        & ANN type                                    & Evaluation                     \\ \hline
\multirow{17}{*}{Classification} & \multirow{2}{*}{3}      & \multirow{2}{*}{2006} & \multirow{2}{*}{\cite{Petushi-2006-LCO}} & \multirow{2}{*}{S. Petushi, et al.}   & 24 slide images,                                        & \multirow{2}{*}{Neural network}             & \multirow{2}{*}{Acc = $90\%$}    \\
                                 &                         &                       &                                     &                                       & H\&E staining                                           &                                             &                                \\ \cline{2-8} 
                                 & \multirow{3}{*}{2}      & \multirow{3}{*}{2010} & \multirow{3}{*}{\cite{osareh-2010-MLTD}} & \multirow{3}{*}{A. Osareh, et al.}    & Dataset 1: 692 specimens of fine needle aspirates of    & \multirow{3}{*}{PNN}                        & \multirow{3}{*}{Acc = $98.8\%$}  \\
                                 &                         &                       &                                     &                                       & breast lumps,                                           &                                             &                                \\
                                 &                         &                       &                                     &                                       & Dataset 2: 295 microarrays                              &                                             &                                \\ \cline{2-8} 
                                 & \multirow{3}{*}{4}      & \multirow{3}{*}{2011} & \multirow{3}{*}{\cite{Singh-2011-BCD}}   & \multirow{3}{*}{S. Singh, et al.}     & 1080 images,                                            & Feed forward                                & \multirow{3}{*}{Acc = $95\%$}    \\
                                 &                         &                       &                                     &                                       & H\&E staining,                                          & back propagation                            &                                \\
                                 &                         &                       &                                     &                                       & ( 1080 for training, 387 for validation, 387 for test ) & neural network                              &                                \\ \cline{2-8} 
                                 & \multirow{3}{*}{3}      & 2011                  & \cite{Zhang-2011-BCC}                    & \multirow{3}{*}{Y. Zhang, et al.}     & 361 images,                                             & \multirow{3}{*}{MLP}                        & \multirow{3}{*}{Acc = $95.22\%$} \\ \cline{3-4}
                                 &                         & 2013                  & \cite{Zhang-2013-BCD}                    &                                       & H\&E staining,                                          &                                             &                                \\ \cline{3-4}
                                 &                         & 2013                  & \cite{Zhang-2013-BCH}                    &                                       & ( $760 \times 570$ )                                             &                                             &                                \\ \cline{2-8} 
                                 & \multirow{3}{*}{3}      & \multirow{3}{*}{2013} & \multirow{3}{*}{\cite{Loukas-2013-BCC}}  & \multirow{3}{*}{C. Loukas, et al.}    & 65 regions of interests                                 & \multirow{3}{*}{PNN}                        & \multirow{3}{*}{Acc = $87\%$}    \\
                                 &                         &                       &                                     &                                       & H\&E staining,                                          &                                             &                                \\
                                 &                         &                       &                                     &                                       & ( 20 grade I, 20 grade II, 25 grade III )               &                                             &                                \\ \cline{2-8} 
                                 & \multirow{3}{*}{2}      & \multirow{3}{*}{2017} & \multirow{3}{*}{\cite{shukla-2017-CHIB}} & \multirow{3}{*}{K. Shukla, et al.} & 70 images,                                              & \multirow{3}{*}{MLP}                        & Acc = $80\%$,                    \\
                                 &                         &                       &                                     &                                       & H\&E staining,                                          &                                             & Sn = $82.9\%$,                    \\
                                 &                         &                       &                                     &                                       & ( 35 non-cancerous and 35 cancerous )                   &                                             & AUC = $89.2\%$                   \\ \hline
\multirow{5}{*}{Segmentation}    & \multirow{2}{*}{Nuclei} & \multirow{2}{*}{2013} & \multirow{2}{*}{\cite{Kowal-2013-CDO}}   & \multirow{2}{*}{M. Kowal, et al.}     & 500 cytological samples,                                & \multirow{2}{*}{Competitive neural network} & \multirow{2}{*}{Acc = $98.7\%$}  \\
                                 &                         &                       &                                     &                                       & H\&E staining                                           &                                             &                                \\ \cline{2-8} 
                                 & \multirow{3}{*}{Nuclei} & \multirow{3}{*}{2013} & \multirow{3}{*}{\cite{Mouelhi-2013-ASS}} & \multirow{3}{*}{A. Mouelhi, et al.}   & 24 microscopic images                                   & \multirow{3}{*}{MLP}                        & \multirow{3}{*}{Acc = $95.5\%$}  \\
                                 &                         &                       &                                     &                                       & IHC staining,                                           &                                             &                                \\
                                 &                         &                       &                                     &                                       & ( $2048 \times 1360$ )                                         &                                             &                                \\ \hline
\end{tabular}
\label{table1}
\end{table*}

\section{BHIA Using Deep Neural Networks}
\label{s:deep} 

In the analysis of breast histopathology images based on deep neural networks, some publicly available datasets are frequently applied. As shown in \tablename~\ref{table2}, we provide detailed information and download links for the datasets mentioned in our review.

\begin{table*}[htbp!]
\renewcommand\arraystretch{2}
\setlength{\tabcolsep}{1 pt}
\centering
\scriptsize\caption{Popular publicly available breast histopathology image dataset. The fourth column ``Detail'', shows the number of classes.}
\begin{tabular}{|c|c|c|c|c|c|c|}
\hline
Datasets                           & Year                  & Staining              & Detail                            & Magnification                                    & Dataset size                                                                 & Website                               \\ \hline
ICPR 2012                           & 2012                  & H\&E                  & \textbackslash{}                  & $40\times$                                       & 50 images corresponding to 50 high-power fields in 5 different biopsy slides & Closed                                 \\ \hline
\multirow{2}{*}{IDC}               & \multirow{2}{*}{2014} & \multirow{2}{*}{H\&E} & \multirow{2}{*}{\textbackslash{}} & \multirow{2}{*}{$40\times$}                      & 277,524 patches are from 162 IDC breast cancer histopathological slides      & \multirow{2}{*}{~\cite{IDC-2014}}     \\
                                   &                       &                       &                                   &                                                  & (198,738 IDC negative, 78,786 IDC positive)                                  &                                       \\ \hline
BreaKHis                           & 2015                  & H\&E                  & 4                           & $40\times$, $100\times$, $200\times$,$400\times$ & 7,909 histopathology images                                                   & ~\cite{BreakHis-2015}                 \\ \hline
Bioimaging 2015 breast         & \multirow{2}{*}{2015} & \multirow{2}{*}{H\&E} & \multirow{2}{*}{4}          & \multirow{2}{*}{$200\times$}                     & 249 images for training, 20 image for testing                                & \multirow{2}{*}{~\cite{BCH2015-2015}} \\
histology classification challenge &                       &                       &                                   &                                                  & and an extended testset of 16 images                                         &                                       \\ \hline
TUPAC 2016                         & 2016                  & H\&E                  & \textbackslash{}                  & $40\times$                                       & 500 for training and 321 for testing breast cancer histopathology WSIs       & ~\cite{TUPAC2016-2016}                \\ \hline
Camelyon 2016                      & 2016                  & H\&E                  & 2                           & $40\times$,$10\times$,$1\times$                  & 400 WSIs of lymph node                                                       & ~\cite{Camelyon2016-2016}             \\ \hline
Camelyon 2017                      & 2017                  & H\&E                  & \textbackslash{}                  & $40\times$                                       & 200 WSIs of lymph node                                                       & ~\cite{Camelyon2017-2017}             \\ \hline
\multirow{2}{*}{BACH}              & \multirow{2}{*}{2018} & \multirow{2}{*}{H\&E} & \multirow{2}{*}{4}          & \multirow{2}{*}{\textbackslash{}}                & Part A: 400 microscopy images                                                 & \multirow{2}{*}{~\cite{BACH-2018}}    \\ \cline{6-6}
                                   &                       &                       &                                   &                                                  & Part B: 30 whole-slide images                                                 &                                       \\ \hline
\end{tabular}
\label{table2}
\end{table*}

\subsection{Related Works}
\label{ss:deep:related}

In this section, we group related work according to the applied datasets. Then, we summarize the motivation, contribution, methods, and results of each paper in chronological order.

\paragraph{\textbf{``BreaKHis'' Tasks: }} 

In 2015, BreaKHis dataset was released in~\cite{Spanhol-2015-ADF}. This dataset is composed of 7,909 histopathological images from 82 clinical breast cancer patients. All the histopathological images of breast cancer are three channel RGB micrographs with a size of $700 \times$ 460 pixels. Since objective lenses of different multiples are used in collecting these histopathological images of breast cancer, the entire dataset comprises four different sub-datasets, namely $40\times$, $100\times$, $200\times$, and $400\times$. All of these sub-datasets are grouped into benign and malignant tumors. Based on this dataset, many related works are carried out.

\textit{Related Works of BreaKHis in 2016:} 

\begin{figure}[htbp!]
\centering
\includegraphics[width=0.6\linewidth]{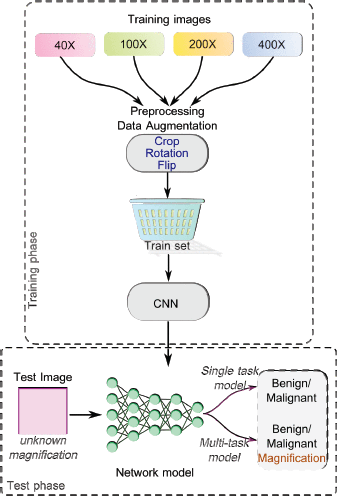}
\caption{Schematic presentation for classifying breast histology images in ~\cite{bayramoglu-2016-DLMI}. This figure corresponds to Fig.3 in the original paper.}
\label{fig:workflow21}
\end{figure}

In ~\cite{bayramoglu-2016-DLMI}, the classification of breast cancer histopathological images by a \textit{Convolutional Neural Network} (CNN) independent of magnification is proposed. This paper uses two different architectures: Single-task CNN is used to predict malignancy, while multi-task CNN is used to predict both malignancy and image magnification levels simultaneously. \figurename~\ref{fig:workflow21} is the overall process of this work. Finally, the average recognition rate of the single-task CNN model in the benign/malignant classification task is $83.25\%$. The average recognition rate of the multi-task CNN model in the benign/malignant classification task is $82.13\%$ and the average recognition rate in the magnification estimation task is $80.10\%$.

\textit{Related Works of BreaKHis in 2017:} 

In~\cite{Spanhol-2016-BCH,Spanhol-2017-DFF,Spanhol-2018-ABC}, based on LeNet 
and AlexNet, deep ANN methods are used to classify breast histopathology images in 
the BreaKHis dataset. In the experiment, the dataset is divided into training ($70\%$) 
and testing ($30\%$) sets, and an overall accuracy around $85\%$ is obtained.

In~\cite{Song-2017-AFV}, a transfer learning work is carried out, where an image is first represented by Fisher Vector (FV) encoding of local features extracted using the CNN model pre-trained on ImageNet. Then, a new adaptation layer is designed to fine-tune the whole deep learning structure. Finally, an accuracy around 
$87\%$ is achieved on $30\%$ testing images. 
Similarly, in~\cite{Zhi-2017-UTL}, another transfer learning strategy is applied to the same task, and achieves an overall accuracy around $90\%$. 

In~\cite{Nejad-2017-COH}, a deep learning structure with a single convolutional layer is proposed for classification task, which obtains an accuracy of $77.5\%$. 
In contrast, in~\cite{Li-2017-UDL}, a deep learning model with multi-layer CNNs is built up, 
and obtains an accuracy up to $90\%$.
Furthermore, in~\cite{Han-2017-BCM}, a CNN model, namely the ``Class Structure-based Deep CNN'' (CSDCNN), is proposed to represent the spatial information within a deep CNN.

In~\cite{das-2017-CHWU}, a DCNN based whole-slide histopathology classifier is presented. First, the posterior estimate of each view at a specific magnification is obtained from CNN at a specific magnification. Then the posterior estimate across random multi-views at multi-magnification is voting filtered to provide a slide level diagnosis. Finally, the experiment uses a patient-level 5-folded cross-validation and achieves an average accuracy of $94.67\%$, sensitivity of $96\%$, specificity of $92\%$ and F-score of $96.24\%$.

In~\cite{wei-2017-DLMB}, a new method for breast cancer histopathological image classification based on DCNNs is proposed, called the BiCNN model, for two-class classification problems on pathological images. The BiCNN has more depth, more width and more complex architecture, which has little parameters and reliable performance. In the experiment, the average recognition rate for patient level is achieved as $97\%$.

\textit{Related Works of BreaKHis in 2018:} 

In~\cite{Motlagh-2018-BCH}, different ResNet structures are tested and compared for this task. The ResNet-V1-152 model obtains the best performance with an overall accuracy of $99.6\%$ after 3000 epochs. 
Similarly, in~\cite{Mehra-2018-BCH}, the effectiveness of three well recognized pre-trained transfer learning models (VGG-16, VGG-19, and ResNet-50 networks) are compared in this task. In the experiment, the VGG-16 with a logistic regression classifier obtains the best performance of $92.6\%$ accuracy. Furthermore, in~\cite{Nawaz-2018-ACO}, Inception-V1, Inception-V2, and ResNet-V1-50 based transfer learning methods are compared, and ResNet-V1-50 obtains the highest accuracy of $95\%$.

In~\cite{Nahid-2018-HBC2}, two restricted Boltzmann machine and back propagation based DCNN models are proposed. Using these two models, the best average accuracy of $87.75\%$ is obtained. Furthermore, in~\cite{Nahid-2018-HBC}, based on CNN and Recurrent Neural Network (RNN) algorithms, a combined deep learning structure is introduced. In this work, unsupervised learning algorithms are first used to segment different tissues into different regions. Then, based on the segmentation result, the proposed deep learning approach is applied to the final classification task. Lastly, an accuracy of $91\%$ is achieved. In addition, in the work of~\cite{Nahid-2018-HBC3}, five DCNN models are built up, considering handcraft features and deep learning features jointly. In the experiment, the second model obtains the best performance of $92.19\%$ accuracy.

In~\cite{Du-2018-BCH}, a classification approach via deep active learning and confidence boosting is introduced, and achieves an overall accuracy of around $90\%$. Similarly, in~\cite{Lee-2018-DLA}, an implemented in-house CNN model is proposed, which combines the advantages of both machine learning features and classical color features. 

In~\cite{Nawaz-2018-MBC}, a DenseNet based CNN model is proposed for classification, including four dense blocks and three transition layers. In the experiment, a $95.4\%$ accuracy is achieved. Similarly, in~\cite{Gandomkar-2018-AFF}, a ResNet based 152 layer deep learning model is built and achieves a correct classification rate of $98.77\%$.

In~\cite{DalalBardou-2018-CBCB}, CNNs are directly compared to classification based on hand-crafted features in binary classification (benign and malignant) and multi-class classification (benign and malignant sub-classes ) of breast cancer histological images. The results show that CNNs outperformed the hand-crafted feature based classifier, where the accuracy reach between $96.15\%$ to $98.33\%$ for the binary classification and $83.31\%$ to $88.23\%$ for the multi-class classification.

In~\cite{KausikDas-2018-MLDC}, a multiple instance learning framework for CNN is proposed. A new pooling layer is proposed that would help to gather most of the informative features from the patches that make up the whole slide, without inter-patch overlap or global slide coverage. In the experiment, at $40\times$, $100\times$, $200\times$ and $400\times$ magnifications, the accuracy is $89.52\%$, $89.06\%$, $88.84\%$ and $87.67\%$, respectively.

In~\cite{Shallu-2018-AMIC}, a new model for automatic classification of breast cancer tissue in histological images by DCNN is proposed, which does not consider the magnification factor of the image. The experimental results on BreaKHis achieved an average accuracy of $85.3\%$.

In~\cite{cascianelli-2018-DRSC}, three dimensionality reduction strategies, including PCA, Gaussian Random Projection (GRP) and Correlation based Feature Selection (CBFS), are applied to CNN-based features to classify histological images of breast cancer. In the experiment, the BreaKHis dataset, Epistroma dataset, and the Multi-class Kather's dataset are tested. Finally, BreaKHis dataset at $40\times$, $100\times$, $200\times$ and $400\times$ magnifications, the accuracy is $87.0\%$, $85.2\%$, $85.0\%$ and $81.3\%$ , respectively. On the Epistroma dataset, the accuracy is $94.7\%$. On Multi-class Kather's dataset, the accuracy of $84.0\%$ is obtained.

\textit{Related Works of BreaKHis in 2019:} 

In~\cite{xu-2019-LICA}, a novel framework based on the hybrid attention mechanism is proposed to classify breast cancer histopathology images. This framework could automatically find useful regions from raw images, and thus does not have to resize the raw images for the network to prevent information loss. At four different magnifications, the average accuracy is about $96\%$ while only $15\%$ of raw pixels are used.

\begin{figure}[htbp!]
\centering
\includegraphics[width=1\linewidth]{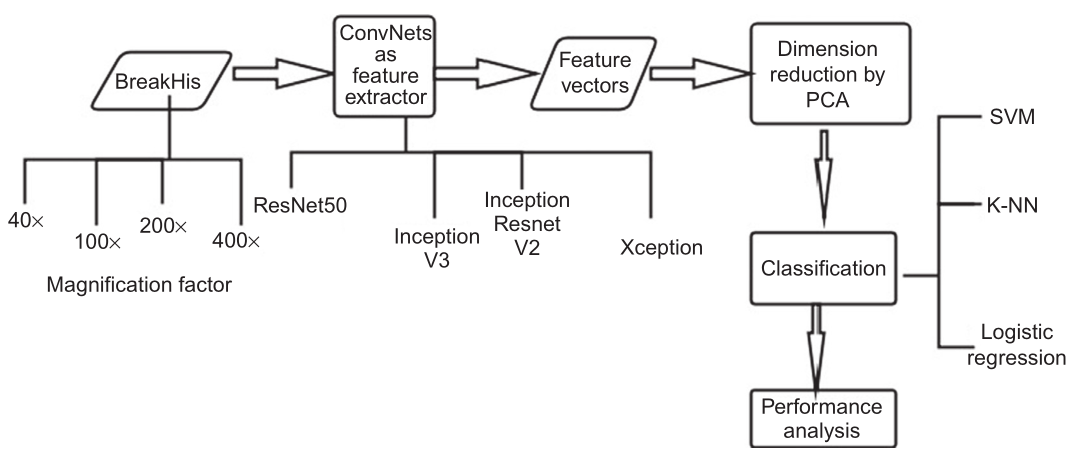}
\caption{Overall structure of the proposed model in ~\cite{bhuiyan-2019-TLSC}. This figure corresponds to Fig.4.6 in the original paper.}
\label{fig:w8}
\end{figure}

In~\cite{bhuiyan-2019-TLSC}, a transfer learning and supervised classifier based prediction model for breast cancer is proposed. As can be seen from \figurename~\ref{fig:w8}, four pre-trained ConvNets are used for transfer learning to extract image features, and then PCA is applied to the feature vectors to reduce feature dimension. Finally, SVM, KNN, and Logistic regression are respectively used to classify images. In the experiment, the best results are obtained on $40 \times $. The ResNet-50 with SVM classifier has the maximum accuracy of $96.24\%$ and the best recall value of $100\%$. On the other hand, with Inception ResNet-V2, SVM gives the highest precision of $96.55\%$.

In~\cite{xie-2019-DLBA}, Inception-V3 and Inception-ResNet-V2 are trained using transfer learning techniques for binary and multi-class classification of breast cancer histopathological images. The results show that the Inception-ResNet-V2 network achieves the best results at $40\times$ magnification: In the binary classification task, the image level accuracy is $97.90\%$, and in the multi-classification task, the image level accuracy is $92.07\%$.

In~\cite{YunJiang-2019-BCH}, a breast cancer histopathology image classification network (BHCNet) is designed. BHCNet includes one plain convolutional layer, three SE-ResNet~\cite{hu-2018-SEN} blocks, and one fully connected layer. Each SE-ResNet block is stacked by $N$ small SE-ResNet modules, which is denoted as BHCNet-$N$. In the results, the BHCNet-3 achieves the accuracy between $98.87\%$ and $99.34\%$ for the binary classification and the BHCNet-6 achieves the accuracy between $90.66\%$ and $93.81\%$ for the multi-class classification.

In~\cite{thuy-2019-FDLT}, deep learning, transfer learning and \textit{Generative Antagonistic Network} (GAN) are combined to improve the accuracy of breast cancer classification on a limited training dataset. First, the fine-tuned VGG-16 and VGG-19 are used to extract features and sent to CNN for classification. In addition, StyleGAN~\cite{karras-2019-ASGA} and Pix2Pix~\cite{isola-2017-ITCA}, two GAN models, are applied to generate 4,800 and 2,912 fake images, respectively. In the experiment, the proposed method is evaluated on the BreaKHis dataset and two generated datasets from BreaKHis by GAN. The experiments show that GAN images created much noise and affected classification accuracy. Finally, the best result is obtained in BreaKHis dataset. The accuracy is $98.1\%$ in the binary classification.

\begin{figure}[htbp!]
\centering
 \includegraphics[width=1\linewidth, height=4cm]{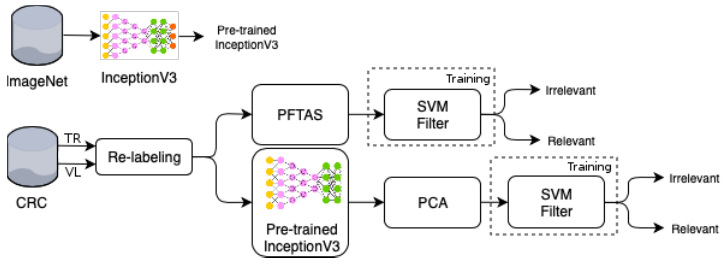}
\caption{The idea of building the filter in~\cite{Matos-2019-DTLB}. PFTAS - Parameter Free Threshold Adjacency Statistics (hand-crafted features), CRC - colorectal cancer dataset, TR - training set, VL - validation set. This figure corresponds to Fig.1 in the original paper.}
\label{fig:w14-1}
\end{figure}

\begin{figure}[H]
\centering
 \includegraphics[width=1\linewidth, height=2.5cm]{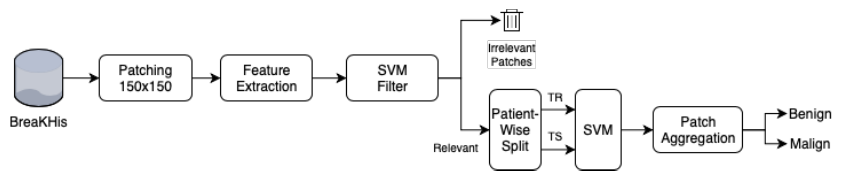}
\caption{An overview of building the classifier in~\cite{Matos-2019-DTLB}: Patching, feature extraction (PFTAS or Inception-v3 + PCA), filtering by SVM, patient-wise splitting of
relevant patches into training (TR) and test (TS) using the pre-defined folds, patch classification and aggregation using majority vote or sum rule. This figure corresponds to Fig.5 in the original paper.}
\label{fig:w14-2}
\end{figure}

In~\cite{Matos-2019-DTLB}, a method of classifying breast cancer histopathologic images based on double transfer learning is proposed. This method can be divided into two steps. In the first step, as shown in \figurename~\ref{fig:w14-1}, in order to improve the quality of the dataset before training the classifier using the BreaKHis dataset, an SVM is trained to classify relevant and irrelevant images in histopathological images, and then it is used as a filter to eliminate irrelevant images from the BreaKHis dataset. In the second step, as shown in \figurename~\ref{fig:w14-2}, another SVM is trained to classify benign and malignant. Both steps use transfer learning (Inception-v3 CNN pre-trained with ImageNet dataset). The best classification accuracy can be obtained by Inception-v3 + filter (Inception-v3 is used to extract features, and filtering refers to the removal of irrelevant images) method at $100\times$ and $200\times$ magnifications of $91\%$ and $89\%$, respectively.

\textit{Related Works of BreaKHis in 2020:} 

In ~\cite{saxenapre-2020-PCNN}, a novel feature extraction method is proposed for the classification of breast histopathological images. First, the images are divided into small pieces that are not overlapped. Then, pre-trained CNNs (10 models in total) are used for feature extraction. Finally, an SVM is applied as a classifier. In the experiment, the best patient-level accuracy is obtained by the AlexNet-SVM model. At $40\times$, $100\times$, $200\times$ and $400\times$ magnifications, the accuracy is $89.46\%$, $92.61\%$, $93.92\%$, and $89.78\%$, respectively.

In~\cite{gour-2020-RLBC}, a ResHist model is designed, which is a residual learning-based 152-layered CNN to classify the histopathological images of breast cancer. In the experiment, histopathological images are first augmented and the ResHist model is trained end-to-end on the augmented dataset in a supervised learning manner. Finally, images are classified into benign and malignant categories by using the trained ResHist model. The result shows that the ResHist model achieves a best accuracy of $92.52\%$ and a F1-score of $93.45\%$. In addition, in order to study the discrimination ability of deep features of ResHist model, the extracted feature vectors are fed into KNN, random forest~\cite{breiman-2001-RF}, quadratic discriminant analysis, and SVM classifiers. Among them, when the deep features are fed back to the SVM classifier, the best accuracy of $92.46\%$ is achieved.

\paragraph{\textbf{``Camelyon'' Tasks: }} 
``Camelyon Grand Challenge'' is a task to evaluate computational systems for the automated detection of metastatic breast cancer in WSIs of sentinel lymph node biopsies.

\textit{Related Works of Camelyon 2016:} 

In~\cite{Liu-2017-DCM}, a DCNN is built for this task, and achieves an AUC of $97\%$. In~\cite{Wang-2016-DLF}, a GoogLeNet based deep learning method is introduced, where 270 images are used for training, and 130 are used for testing. Lastly, an AUC of $92.5\%$ is obtained. With the same experimental setting, in~\cite{BenTaieb-2018-PCW}, a recurrent visual attention model is proposed, which includes three primary components composed of dense or convolutional layers to describe the information flow between components within one time-step. Finally, a $96\%$ AUC is achieved.

In~\cite{lin-2018-SAFD}, a fast and dense screening framework (ScanNet) for detecting metastatic breast cancer from WSIs is proposed. ScanNet is implemented based on the VGG-16 network by changing the last three fully connected layers to fully convolutional layers. In the result, Free Response Operating Characteristic (FROC) of 0.8533 and AUC of $98.75\%$ are obtained.

In~\cite{HPang-2018-UTLB}, Multiple Magnification Feature Embedding (MMFE) is introduced, which is an approach using transfer learning to detect breast cancer from digital pathology images without network training. The main idea of the MMFE method is to simulate the daily diagnosis process of a medical doctor. First, a low-resolution image is observed to identify suspicious areas. Then, it is switched to a high-resolution image for further confirmation. Experiments show that this approach can greatly improve the training and prediction speed of the model without reducing the performance of the model.

In~\cite{Bejnordi-2017-DAO}, a summary of the Camelyon 2016 shows that: 25 of 32 submitted algorithms are deep learning based methods, and 19 top-performing algorithms are all DCNN approaches.

\textit{Related Works of Camelyon 2017:} 

In the Camelyon 2017~\cite{Chervony-2017-FCO}, in order to detect four types of breast cancer from the WSIs, a deep learning architecture is proposed with limited computational resources. In this work, two CNNs are applied in a cascade, followed by local maxima extraction and SVM classification of local maxima regions. In the experiment, 300 images are used for training, 200 are used for validation, 500 are for test, and an accuracy of $92\%$ is finally achieved.

\paragraph{\textbf{``BACH'' Tasks: }} 

The Grand Challenge on BreAst Cancer Histology images (BACH) is co-organized with the 15th International Conference on Image Analysis and Recognition (ICIAR 2018)~\cite{aresta-2019-BGCB}. There are two goals in this challenge. Part A of the challenge consists of automatically classifying H\&E stained breast histology microscopy images in four classes: Normal, benign, \textit{in-situ} carcinoma and invasive carcinoma. The data in part A is composed of 400 H\&E stained breast histology images. All images are of equal dimensions ($2048 \times 1536$ pixels). Part B consists of performing pixel-wise labeling of WSIs in the same four classes as Part A. The data in part B is composed of 20 WSIs of very large size. Each WSI can have multiple normal, benign, \textit{in-situ}  carcinoma and invasive carcinoma regions.

In order to classify four types of breast cancer histopathology images, an Inception-V3 based deep learning model is introduced in~\cite{Golatkar-2018-COB}. In the experiment, 300 images are used for training, and 100 are used for testing. Finally, an average accuracy of $85\%$ is achieved.

A two-stage CNN model is also proposed in~\cite{Nazeri-2018-TCN}, where the first stage is for pixel-level classification, and the second stage is for image-level classification. In the experiment, an overall accuracy around $94\%$ is obtained.
Similarly, in~\cite{Kiambe-2018-BHI}, a two-stage classification approach is proposed. In the first stage, an AlexNet based feature extraction is applied. In the second stage, three different classifiers are used. In the experiment, an SVM classifier achieves the best result ($99.84\%$ accuracy). 
Similarly, in~\cite{Ranjan-2018-HAF}, the AlexNet is also applied as a basic model to build a hierarchical classification model, and an accuracy of $95\%$ is obtained.

In~\cite{Vesal-2018-CBCH}, a transfer learning-based approach for the classification of breast cancer histology images is presented. Inception-V3 and ResNet-50 CNNs, both pre-trained on the ImageNet database, are used. In the experiment, the Inception-V3 network achieves an average test accuracy of $97.08\%$ for four classes, marginally outperforming the ResNet-50 network, which achieves an average accuracy of $96.66\%$.

In~\cite{Ferreira-2018-CBCH}, an Inception ResNet-V2 is proposed to classify histological images of breast cancer through transfer learning, fine-tuning, and data augmentation. Out of 100 images in each class, 70, 20, and 10 images are randomly selected for training, testing, and validation. The final results show that the accuracy of the test set is $90\%$ and the loss is 0.59, while the accuracy of the validation set is $93\%$ and the loss is 0.23.

\begin{figure}[htbp!]
\centering
\includegraphics[width=1\linewidth, height=4cm]{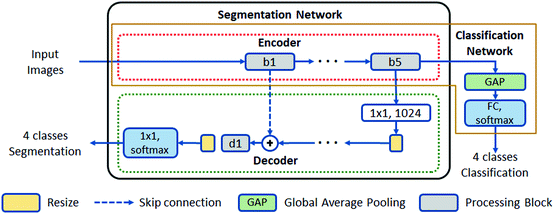}
\caption{Overall structure of the proposed model in ~\cite{Vu-2018-MMBH}. A classification network consists of an encoder and two processing layers. A segmentation network contains an encoder and decoder. This figure corresponds to Fig.1 in the original paper.}
\label{fig:w27}
\end{figure}

In~\cite{Vu-2018-MMBH}, a deep learning approach for analyzing breast histology images at both micro-level (patch-based image classification) and macro-level (WSI segmentation) is proposed. The approach contains two networks, as shown in \figurename~\ref{fig:w27}, which share the architecture and weights, especially the encoder in the network, to improve the utility of the trained network and available dataset. In the result, for patch classification, $71\%$ and $65\%$ accuracy are obtained on the training and test datasets, respectively. In terms of segmentation, an overall score of 0.7343 and 0.4945 are obtained on the training and test data sets, respectively.

In~\cite{MK-2018-ABCH}, approaches for the classification of microscopic images as well as the segmentation of WSIs are presented. In both parts of the challenge, data preparation, scale selection, and augmentation are firstly carried out. In part B of the challenge, additional data are added. Finally, network training is conducted. The densenet-161 architecture with pre-training on ImageNet is selected and repeatedly trained in the expansion training set. The results show that a $96.24\%$ accuracy is achieved.

In~\cite{Wang-2018-BCMI}, a method for classification of breast cancer histopathology images based on deep learning is proposed. The effects of various preprocessing methods are compared, and the classification results of CNN and CNN with SVM are also compared. Finally, an accuracy of $83\%$ is obtained in part A task.

In~\cite{Awan-2018-CLUT}, a context-aware network for automated classification of breast cancer histopathological images is proposed.  The method mainly includes two steps: First, the activation feature of a trained ResNet is used to classify the non-overlapping patches. Then, an SVM classifier is trained to classify the patches of overlapping blocks. Finally, the majority-voting method is used for image-wise classification. As a result, an average accuracy of $83\%$ is obtained.

In~\cite{Cao-2018-IPTL}, six different feature extractors are compared: Hand-crafted features, ResNet-18, ResNeXt, NASNet-A, ResNet-152 and VGG-16. The result shows that the pre-trained deep learning network on ImageNet has better performance than the popular hand-crafted features used for breast cancer histology images. Finally, the integration method based on random forest dissimilarity is used to combine hand-crafted features with five deep learning feature groups, and an average accuracy of $87.1\%$ is obtained.

\begin{figure}[H]
\centering
\includegraphics[width=1\linewidth, height=4cm]{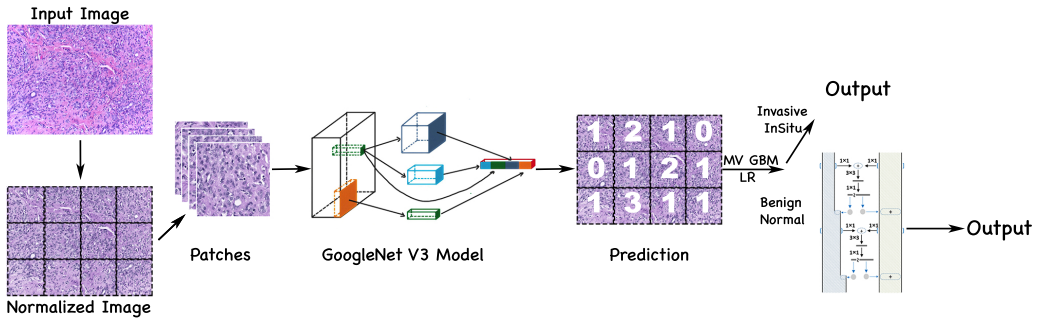}
\caption{An overview of the proposed framework in ~\cite{Vang-2018-DLFM}. This figure corresponds to Fig.2 in the original paper.}
\label{fig:w9}
\end{figure}

In~\cite{Vang-2018-DLFM}, a deep learning framework for multi-class breast cancer image classification is presented. The framework of the approach is illustrated in \figurename~\ref{fig:w9}. First, Inception-V3 is used for patch-wise classification. Then, the patch-wise predictions are passed through an ensemble fusion framework involving majority voting, a Gradient Boosting Machine (GBM), and logistic regression to obtain the image-wise prediction. Finally, an average accuracy of $87.5\%$ is obtained.

In~\cite{yan-2019-BCHI}, a new hybrid convolutional and recurrent deep neural network for breast cancer histopathological image classification is proposed. First, a fine-tuned Inception-V3 is used for feature extraction for each image patch. Then, the feature vectors are input into a 4-layer Bidirectional Long Short-Term Memory network (BLSTM) ~\cite{schuster-1997-BRNN} for feature fusion. Finally, a complete image-wise classification is carried out. In the experiment, an average accuracy of $91.3\%$ is obtained in image-wise. Notably, a new dataset containing 3,771 histopathological images of breast cancer is published in this paper. It covers as many different subclasses spanning different age groups as possible. The dataset is publicly available at~\cite{Yan-2019}.

\begin{figure}[htbp!]
\centering
\includegraphics[width=1\linewidth]{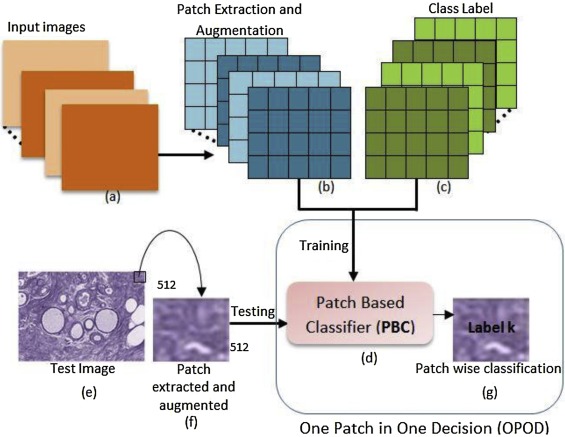}
\caption{Overall structure of OPOD technology in ~\cite{K.Roy-2019-PBSC}. It is worth noting that (g) Patch label prediction by the trained patch based classifier (PBC) where $k$ $\epsilon$ $\lbrace0, 1, 2, 3\rbrace$ . This figure corresponds to Fig.7 in the original paper.}
\label{fig:w22-1}
\end{figure}

\begin{figure}[htbp!]
\centering
\includegraphics[width=1\linewidth]{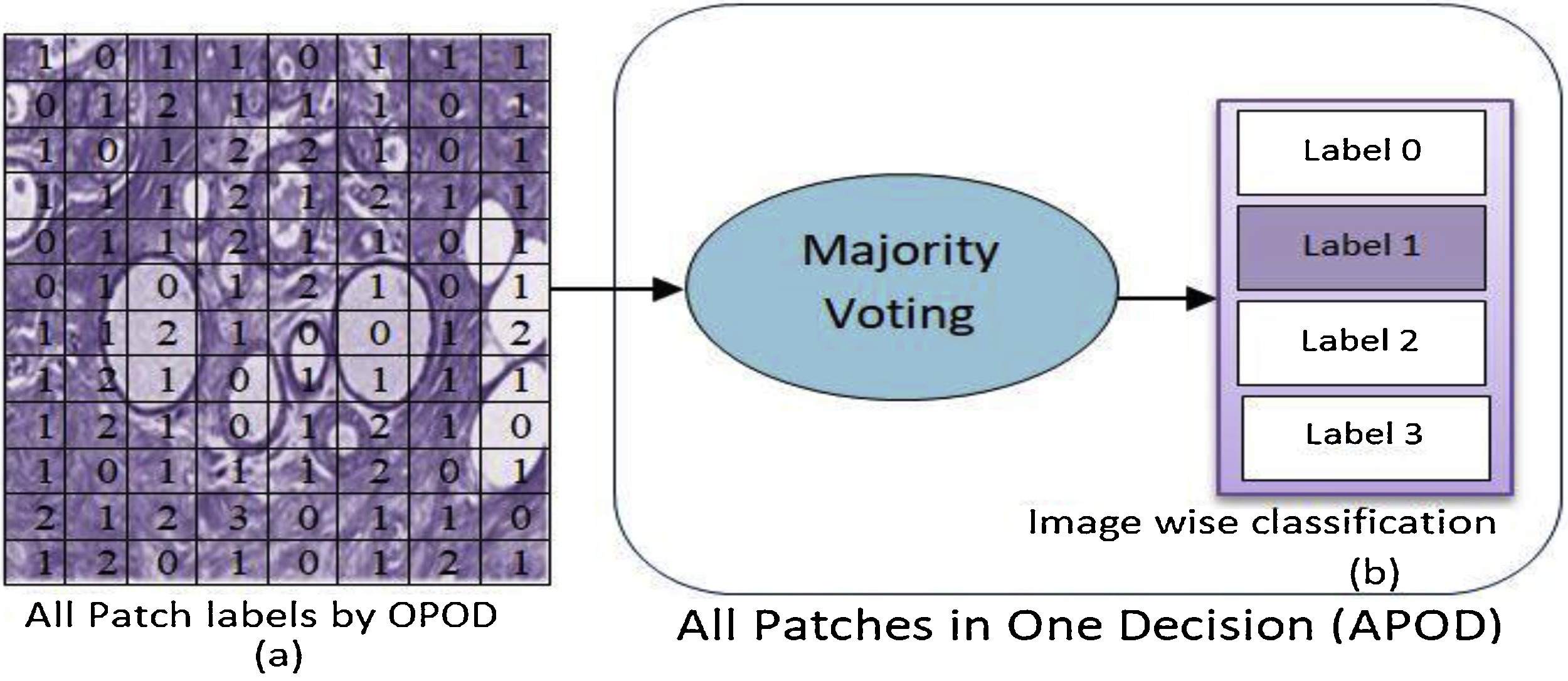}
\caption{Overall structure of the APOD technology in ~\cite{K.Roy-2019-PBSC}. (a) Patch labels of an image predicted by OPOD technique. (b) Image label prediction based on patch label majority voting by proposed APOD technique. This figure corresponds to Fig.8 in the original paper.}
\label{fig:w22-2}
\end{figure}

In~\cite{K.Roy-2019-PBSC}, a patch-based classifier (PBC) using CNN for automatic classification of histopathological breast images is proposed. The proposed classification system works in two different modes: One patch in one decision (OPOD) and all patches in one decision (APOD). The flowchart of OPOD technology for patch classification is shown in \figurename~\ref{fig:w22-1}. OPOD is mainly responsible for predicting the class labels of each patch extracted from the pre-processed histopathological images. As shown in \figurename~\ref{fig:w22-2}, APOD technology classifies images by majority voting for each patch predicted by OPOD technology. In the test set, APOD technology achieves accuracy of $90\%$ in 4-class classification and $92.5\%$ in 2-class classification.

In~\cite{kassani-2019-BCDT}, a method for the diagnosis of breast cancer histopathology images based on transfer learning and global pooling is proposed. Five DCNN architectures are used as feature extractors, namely Inception-V3, InceptionResNet-V2, Xception, VGG-16, and VGG-19. The experimental results show that the network structure based on the pre-trained Xception model is better than all other DCNN structures in average classification accuracy, reaching $92.50\%$.

In~\cite{kausar-2019-MRBH}, a DCNN model with Haar wavelet decomposed images is introduced to classify breast histopathological images. Haar wavelet transform is used to decompose the input high-resolution histopathological image to a small size. In this way, the convolution time of deep CNNs and computational resources can be greatly reduced without any performance downgrade. Meanwhile, this downsampling process also eliminates the need to extract small patches, which helps to improve the accuracy. In the experiment, BACH datasets and BreaKHis datasets are tested. Finally, $98.2\%$ and $96.85\%$ accuracies are obtained on the BACH datasets for both 4-class and 2-class and BreaKHis datasets for multi-class, respectively.

\paragraph{\textbf{``ICPR 2012'' Tasks: }} 

In the 2012 International Conference on Pattern Recognition (ICPR), a ``mitotic figure recognition contest'' is released. The dataset is made up of 50 High Power Fields (HPF) coming from 5 different slides scanned by three different types of equipment at $40\times$ magnification. An HPF has a size of $512 \mu m \times 512 \mu m$. These 50 HPFs contain a total of 326 mitotic cells on images of both scanners and 322 mitotic cells on the multispectral microscope~\cite{roux-2013-MDBC}.

In~\cite{Malona-2012-MFR,Malon-2013-COM}, manually designed color, texture, and shape features are jointly used with the machine learning features extracted by a multi-layer CNN. Finally, this method obtains an F1-scores up to $65.9\%$ on color scanners and $58.9\%$ on multi-spectral scanners.
Similarly, in the work of~\cite{Wang-2014-CEO}, handcrafted features and DCNN features are used in an ensemble learning process together, and an F1-score of $73.5\%$ is obtained.

In~\cite{Ciresan-2013-MDI}, in order to detect the mitosis in a breast histology image, a deep max-pooling CNN is built up, which is trained to classify each pixel in the image into a labeled region. In the experiment, 26 images are used for training, 9 for validation, and 15 for test. Finally, an F1-score of $78.2\%$ is achieved. Furthermore, a similar method is used in the work of~\cite{Veta-2014-BCH2}, 
and an F1-score of $61.1\%$ is obtained.

In~\cite{chen-2016-MDBC}, a novel deep cascade convolutional neural network (CasCNN) is designed to detect mitosis. CasCNN consists of two parts. First, using full CNN, a rough retrieval model is proposed to identify and locate mitotic candidates while maintaining high sensitivity. Then, a fine recognition model based on cross-domain knowledge transfer is proposed to further single out mitoses from the rough model. In the experiment, both the ICPR12 dataset and ICPR14 dataset are used. On the ICPR12 dataset, the precision of $80.4\%$, recall of $72.2\%$ and F1-score of $78.8\%$ are obtained. 
On the ICPR14 dataset, the precision of $46\%$, recall of $50.7\%$ and F1-score of $48.2\%$ are obtained. 

In~\cite{roux-2013-MDBC}, a summary of the ICPR 2012 contest shows that 17 teams submit their results and the IDSIA team gets the best performance. In the work of the IDSIA team, a CNN is trained through ground truth mitosis provided in training dataset, and then the CNN is used to calculate a map of mitosis probabilities on the whole image. Finally, achieving a recall of $70\%$, the accuracy of $89\%$, and the F-measure of $78\%$.

\paragraph{\textbf{``TCUG16'' Tasks: }} 

The Tumor Proliferation Assessment Challenge 2016 (TUPAC16) is the first challenge to predict tumor proliferation scores from WSIs. This challenge is organized in the context of the MICCAI 2016 conference in Athens, Greece. The goal of the challenge is to assess algorithms that predict the tumor proliferation scores from the WSIs. There are two tasks in this challenge. Task 1 is to predict the proliferation score based on mitosis counting. Task 2 is about the prediction of proliferation score based on molecular data. The participants can submit their results for either or both of the tasks.

In~\cite{Wahab-2019-TLBD}, a transfer learning system based DCNN algorithm is suggested for the segmentation and detection of mitoses in breast cancer histopathological images. This system uses two CNNs. A pre-trained CNN is used for segmentation, and Hybrid-CNN is employed for mitotic classification. Finally, in the task of mitosis detection, an F-measure of $71.3\%$ with $76\%$ area under the precision-recall curve is achieved.

In~\cite{ChaoLi-2019-WSMD}, a novel technique for the detection of mitosis by virtue of semantic segmentation is presented, called SegMitos. At the same time, a novel concentric label and concentric loss are proposed, which can train a dense prediction model with weak annotation. The idea of the experiment is as follows. First, the preparation of data and to produce concentric labels. Then to train the SegMitos model. Finally, the trained model is deployed to the testing images of the mitosis dataset. As a result, four datasets (ie: 2012 ICPR MITOSIS dataset, MITOS-ATYPIA-14 dataset, AMIDA13 dataset, and TUPAC16 dataset) are utilized to validate the proposed method. Finally, on the TUPAC16 dataset, an F-score of $66.9\%$ is obtained.

In~\cite{veta-2019-PBTP}, a summary of the TCUG16 shows that: 12 teams submit results for the first task, and 6 teams submit results for the second task. With the exception of one team, all teams use DCNNs as part of the processing pipeline. For the first task, the best performing method achieves a quadratic-weighted Cohen's kappa score of $k$ = 0.567, $95\%$ CI [0.464, 0.671] between the predicted scores and the ground truth. For the second task, the predictions of the top method have a Spearman's correlation coefficient of $r$ = 0.617, $95\%$ CI [0.581, 0.651] with the ground truth.

\paragraph{\textbf{``Bioimaging 2015 Breast Histology Classification Challenge'' Tasks:}}

In 2015, the Bioimaging 2015 Breast Histology Classification Challenge dataset was released in~\cite{Araujo-2017-COB}. This dataset is composed of high-resolution ($2048 \times 1536$ pixels) and H\&E stained breast cancer histology images. All the images are digitized with the same acquisition conditions, with the magnification of $200 \times$ and pixel size of $0.42\mu m \times 0.42\mu m$. There are four different types of histological images of breast cancer in the dataset, namely normal tissue, benign lesion, \textit{in-situ} carcinoma and invasive carcinoma. A total of 285 images are included in the dataset. Of these, 249 images are used for the training set and 36 images are used for the test set. The pictures in the test set are divided into two groups, namely the initial group (20 images with less classification difficulty) and the extended group (16 images with more difficult classification).

In~\cite{Araujo-2017-COB}, a DCNN model is introduced to classify four breast cancer histopathology types in the whole slide image. In the experiment, 249 images are used for training, 20 images are used for testing, and accuracy of $77.8\%$ is obtained.

In~\cite{Mahbod-2018-BCH}, in order to classify different breast cancer types in H\&E stained histopathology images, pre-trained ResNet-50 and ResNet-101 networks are applied with a fine-tuned process and a fusion strategy. In the experiment, the bioimaging 2015 breast histology classification challenge datasets and BACH datasets are tested. Finally, $97.22\%$ and $88.5\%$ accuracies are obtained on the Bioimaging 2015 Breast Histology Classification Challenge dataset and BACH dataset, respectively.

In~\cite{Rakhlin-2018-DCN}, an extended version of the Bioimaging 2015 Breast Histology Classification Challenge dataset is used. This dataset is the same as the original data set in terms of image type, acquisition conditions, and size. The total number of images in this dataset is 400. The pre-trained ResNet-50, Inception-V3, and VGG-16 networks are fused into a deep learning structure and achieve a mean accuracy of $87.2\%$.

\begin{figure}[htbp!]
\centering
\includegraphics[width=1\linewidth, height=6.5cm]{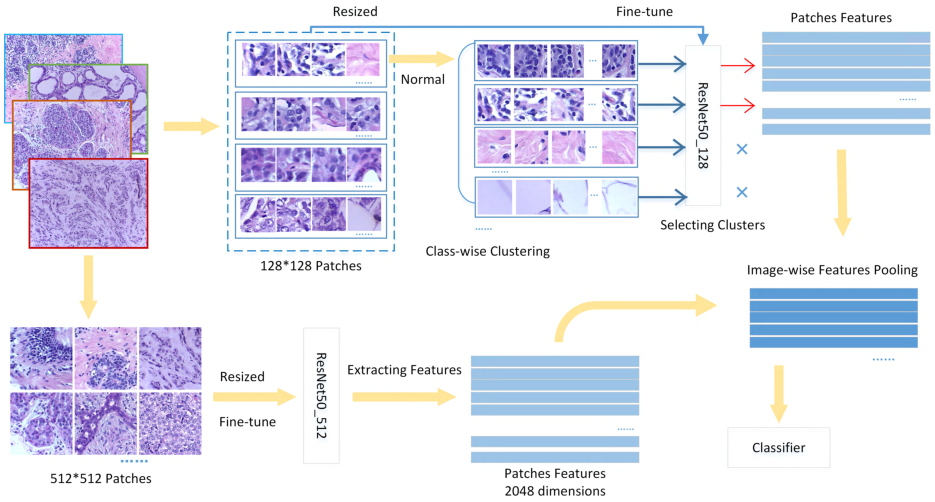}
\caption{A schematic illustration of the proposed framework in ~\cite{YuqianLi-2019-COBC}. This figure corresponds to Fig.3 in the original paper.}
\label{fig:w2}
\end{figure}

In~\cite{YuqianLi-2019-COBC}, an approach based on deep learning for multi-classification of breast histological images is proposed. The framework of the approach is illustrated in \figurename~\ref{fig:w2}. Firstly, two patches of different sizes are extracted from the histological images of breast cancer by sliding window mechanism, including cell-level and tissue-level features. In order to solve the problem of insufficient diagnostic information or label information error in some sampling patches, a batch screening method based on CNN and \textit{k}-means is proposed to select more discriminant patches. Then, ResNet-50 is used as the feature extractor to extract features from the patch and P-norm pooling is used to obtain the final image features. Finally, an SVM is used for the final image classification. The result shows that a $95\%$ accuracy is achieved on the initial test set.

In~\cite{Md-2019-BCCH}, a method for breast cancer classification with the Inception Recurrent Residual Convolutional Neural Network (IRRCNN) model is proposed. The IRRCNN approach is applied for breast cancer classification on two publicly available datasets including the Bioimaging 2015 Breast Histology Classification Challenge dataset and BreaKHis. On the Bioimaging 2015 Breast Histology Classification Challenge dataset, test accuracy of $ 99.05 \% $ and $ 98.59 \% $ is obtained for the binary and multi-class classifications, respectively. On the BreaKHis dataset, the accuracy of $97.95 \pm 1.07\%$ (image-level) and $97.65 \pm 1.20\%$ (patient-level) is obtained for the binary classification, respectively. In addition, the accuracy of $97.57 \pm 0.89\%$ (image-level) and $96.84 \pm 1.13\%$ (patient-level) is obtained for the multi-class classification, respectively. 

In~\cite{Hafiz-2019-COBC}, transfer learning based on AlexNet, GoogleNet, and ResNet is used to classify the histopathological images of breast cancer. The result shows that ResNet has the highest accuracy, achieving $83.60\%$ and $85.0\%$ accuracy at the patch level and image level, respectively.

\paragraph{\textbf{``IDC'' Tasks:}}

Invasive Ductal Carcinoma (IDC) dataset is a publicly available dataset that was first introduced by Cruz-Roa et al.\cite{cruz-2014-ADID}. The dataset contains digital breast cancer histopathological slides from 162 women with IDC. All slides are digitized via a whole-slide scanner at $40\times$ magnification. The dataset contains 277,524 patches of size $50 \times 50$ pixels (198,738 IDC negative and 78,786 IDC positive).

\begin{figure}[htbp!]
\centering
\includegraphics[width=1\linewidth, height=2.8cm]{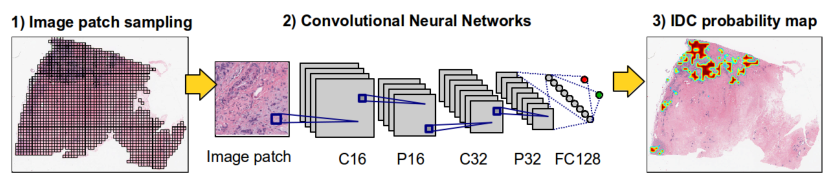}
\caption{Overall framework for automated detection of IDC in WSI using CNN in ~\cite{cruz-2014-ADID}. This figure corresponds to Fig.1 in the original paper.}
\label{fig:w17}
\end{figure}

In~\cite{cruz-2014-ADID}, a deep learning approach for automatic detection and visual analysis of IDC tissue regions in WSIs of breast cancer is presented. The framework of the approach is illustrated in \figurename~\ref{fig:w17}. First, grid-sampling image patches of all the regions containing tissue in WSI. Then, CNN is trained from the sampled patch to predict the probability of patch belonging to IDC. Finally, a probability map is built on the WSI, highlighting patches that have IDC with a probability greater than 0.29. In the experiment, 162 original slices are divided into 3 different subsets: 84 of them are used for training, 29 are used as the validation set and 49 are used for testing. As a result, an F-measure of $71.80\%$ and an accuracy of $84.23\%$ are obtained for automatic detection of IDC regions in WSI.

In~\cite{FP-2019-MBNI}, a new CNN-based model for identifying IDC cells in histopathological slides is proposed. This model, which is derived from the Inception architecture, proposes a multi-level batch normalization module between each convolution step. In the experiment, 94,543 patches are used for training, 31,514 for validation, and 151,465 for testing. Finally, a balanced accuracy of $89\%$ and an F1-score of $90\%$ are obtained.

\paragraph{\textbf{Other Tasks: }} 
\subsubsection{Classification}

In~\cite{Wu-2014-HIC}, a Principal Component Analysis Network (PCANet) is introduced to classify Ductal Carcinoma \emph{\textit{In-Situ}} (DCIS) and Usual Ductal Hyperplasia (UDH) images. In this work, a dataset with 20 DCIS and 31 UDH images are tested, where 10,000 patches are randomly sampled from the training set to learn the models. Finally, an accuracy around $79\%$ is achieved. 

In~\cite{bejnordi-2017-DLAT}, a classification method based on CNN is proposed for the WSIs of breast tissue. In this work, two CNNs are trained. CNN-\uppercase\expandafter{\romannumeral1} is used to classify the WSI into the epithelium, stroma, and fat. CNN-\uppercase\expandafter{\romannumeral2} operates on the stromal regions output by classification of CNN-\uppercase\expandafter{\romannumeral1}, and then classifies the stromal regions as normal stroma or cancer-associated stroma. The dataset contains 646 sections of breast tissue stained with H\&E. In the experiment, 270 images are considered for training, 80 copies are for validation, and 296 are for testing. Finally, an area under Receiver-Operating Characteristic (ROC) of 0.921 is obtained.

In~\cite{Wang-2018-COB}, to distinguish four breast cancer types in histopathological images, a deep learning method is introduced with hierarchical loss and global pooling. In this work, VGG-16 and VGG-19 networks are applied as the basic deep learning structures, and a dataset with 400 images are tested. In the experiment, 280 images are used for training, 60 images are for validation and 60 are for testing. Finally, an average accuracy of around $92\%$ is obtained.

In~\cite{Gecer-2018-DAC}, a work is carried out to classify five diagnostic breast cancer styles in the whole histopathological image. First, a saliency detector performs multi-scale localization of diagnostically relevant regions of interest in the images. Then, a CNN classifies image patches as five types of carcinoma. Lastly, the saliency and classification maps are fused for final categorization. In the experiment, 240 images are used to examine the effectiveness of the proposed method, and a $55\%$ accuracy is finally achieved. The highlight of this work is that 45 pathologists take part in the final evaluation of the test images, and an average accuracy of around $65\%$ is obtained. Hence, the performance of the proposed method is comparable to the performance of the pathologists that practice breast pathology in their daily routines.

In~\cite{HD-2018-IAWD}, an image analysis method is developed that uses deep learning to classify tumor grade, ER status, PAM50 intrinsic subtype, histological subtype, and recurrence risk score (ROR-PT). In the experiment, 571 examples of breast tumors are used for training and 288 are for testing. Finally, it can be distinguished from low-intermediate and high tumor grades ($82\%$ accuracy), ER status ($84\%$ accuracy), Base-like and non-base-like ($77\%$ accuracy), Ductal vs. lobules ($94\%$ accuracy), and high vs. low-medium ROR-PT score ($75\%$ accuracy).

In~\cite{khan-2019-ANDL}, pre-trained CNN architectures (GoogLeNet, VGGNet, and ResNet) are used to extract features from images, and these features are fed to a fully connected layer. The average pool classification is used to classify malignant cells and benign cells. Two breast microscopic image datasets are used: The first is a standard benchmark dataset ~\cite{Spanhol-2015-ADF} and the other is developed locally at LRH hospital in Peshawar, Pakistan. In the experiment, 6000 images are considered to train the architecture and 2000 images are used for testing. Finally, an average classification accuracy of $97.53\%$ is achieved.

In~\cite{Qaiser-2018-HCCA}, Human epidermal growth factor receptor 2 (HER2) Scoring Contest is introduced. HER2 is an important prognostic factor for breast cancer, so the task of automatic HER2 scoring has great clinical significance. The paper shows that a total of 18 submissions from 14 teams are received for evaluation. Among the comprehensive results of all submitted automated methods, 8 of the top 10 teams used CNN-based learning methods. It can be seen that CNN-based learning methods play an important role in HER2 automatic scoring tasks.

\subsubsection{Segmentation}

In~\cite{Pang-2010-CNS}, a CNN based model with three hidden layers is built to segment the breast cancer cell nucleus in histopathological images.  In this work, 58 H\&E stained images are tested, and overall accuracy of around $95\%$ is achieved on both RGB and Lab color spaces. 

In~\cite{su-2015-RSHB}, a fast scanning deep convolutional neural network (fCNN) is proposed for pixel-wise region segmentation. In the work, 92 images are used, which are selected from 20 patients in The Cancer Genome Atlas (TCGA) breast cancer dataset. 75 images are used for training and 17 images are used for testing. In the experiment, it takes only 2.3 seconds to segment an image with size $1000\times1000$ pixels. Also, a mean precision of $91\%$, a mean recall of $89\%$, and a mean F1-score of $85\%$ are achieved.

In~\cite{xu-2016-ADCN}, a DCNN based feature learning is presented to automatically segment or classify epithelial and stromal regions in histopathological images. In this work, colorectal cancer dataset and breast cancer dataset are used separately. The breast cancer dataset consists of 157 H\&E stained images. The data is acquired from two independent cohorts: Netherlands Cancer Institute (NKI: 106) and Vancouver General Hospital (VGH: 51). In the experiment, a superpixel-based scheme is used to over-segment the image into atomic regions. Then, the atomic regions are adjusted to square images of fixed size, and then they are fed back to the DCNN for feature learning. Finally,  F-score of $85.21\%$, $89.10\%$ and accuracy of $84.34\%$, $88.34\%$ are obtained on NKI and VGH, respectively.

\begin{figure}[htbp!]
\centering
\includegraphics[width=1\linewidth, height=5cm]{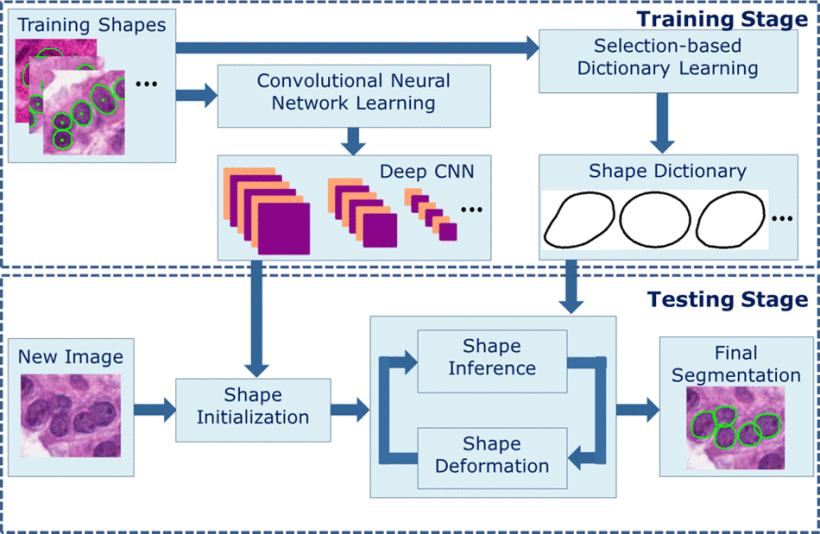}
\caption{The architecture of segmentation framework in~\cite{xing-2016-AALB}. This figure corresponds to Fig.2 in the original paper.}
\label{fig:w20}
\end{figure}

In~\cite{xing-2016-AALB}, an automatic nuclei segmentation technique using DCNN is introduced. \figurename~\ref{fig:w20} depicts a flowchart for the proposed framework. In the training stage, a DCNN model is trained and a kernel shape library based on a selection-based dictionary learning algorithm is obtained. In the test stage, the CNN model is applied to the images to generate probability maps, and then iterative region merging is performed to initialize the shape of each kernel. Then, the proposed kernel segmentation algorithm uses the local repulsive deformation model for shape deformation, and uses the shape priors of the sparse shape model for shape inference. Finally, an accuracy of $92\%$ is achieved on the breast cancer dataset.

\begin{figure}[htbp!]
\centering
\includegraphics[width=1\linewidth, height=4cm]{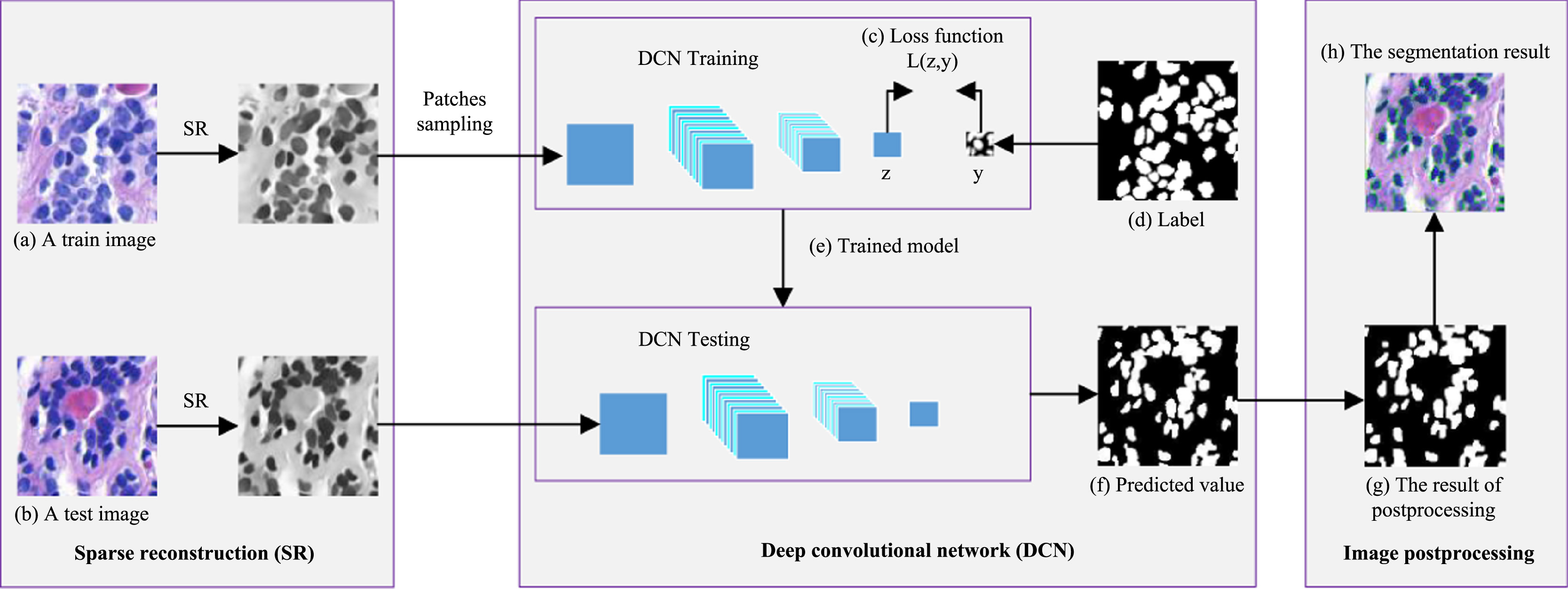}
\caption{An overview of the method proposed in ~\cite{pan-2017-ASNP}. This figure corresponds to Fig.1 in the original paper.}
\label{fig:workflow47}
\end{figure}

In~\cite{pan-2017-ASNP}, an automated nuclei segmentation method is proposed. \figurename~\ref{fig:workflow47} is the overall process of nuclei segmentation. This process can be divided into three main stages. First, the Sparse Reconstruction (SR) method is used to roughly remove the background and highlight the nuclei of the pathological image. Then the gradient descent technique is used to train the DCN cascade of the multi-layer convolutional network in order to effectively segment the nucleus from the background. At this stage, the patch and its corresponding label are randomly extracted from pathological images and input into the training network. Finally, morphological operation and prior knowledge are introduced to improve segmentation performance and reduce errors. In this work, the pixel segmentation accuracy of $92.45\%$ and the F1-measure of $83.93\%$ are obtained.

In~\cite{naylor-2017-NSHI}, a method of nuclear segmentation in histopathological images based on deep learning and mathematical morphology is proposed. In addition, an image dataset containing 33 images with 2754 annotated cells is provided. This dataset can be obtained at~\cite{NPDa-2017-DATA}. In this work, a set of manual annotation images is trained on a deep neural network and the posterior probability map is processed to achieve joint segmentation of the nuclei. Finally, the accuracy of $95.4\%$, recall of $77.3\%$, precision of $86.4\%$ and F1-score of $80.5\%$ are achieved.

In~\cite{CuiYuxin-2018-ADLA}, an advanced supervised full CNN method for nuclear separation in histopathological images is proposed. First, a histopathological image is normalized to the same color space. Then a complete image is split into overlapping small blocks. The proposed nuclear boundary model is used to detect the nucleus and boundary on each plaque, and all the predictions are seamlessly recombined. Finally, fast and parameterless post-processing is applied to generate the kernel segmentation results. In the experiment, multiple organ H\&E stained image dataset (MOD)~\cite{Kumar-2017-ADAT}, breast cancer histopathology image dataset (BCD) and breast cancer image dataset (BNS)~\cite{naylor-2017-NSHI} are used. Finally, an image with a size of $1000 \times 1000$ pixels can be segmented in less than 5 seconds.

In~\cite{Sonia-2019-DACS}, an automatic end-to-end framework using deep neural networks for tissue-level segmentation is proposed. In this work, a new dataset of WSIs with different subtypes of breast cancer, consisting of 11 WSIs fully annotated, is tested. Finally, the results of U-Net, SegNet, FCN, and DeepLab are evaluated by using pixel-by-pixel indexes, with the \textit{Dice Coefficient} (DC) values of 0.86, 0.87, 0.86, and 0.86, respectively.

In~\cite{priego-2020-ASWS}, an automatic WSIs segmentation method based on DCNN is proposed. The method is effective regardless of the texture features in malignant tumors. In addition, the framework is available online and can be used as a computer assisted diagnosis tool by pathologists. In this work, training and testing is performed using 12 breast cancer WSIs stained with H\&E. Finally, the estimated segmentation accuracy is $95.62\%$.

\subsubsection{Detection}
\begin{figure}[htbp!]
  \centering
  \subfigure[]{
    \label{fig:subfig:SSA-seg01} 
    \includegraphics[width=0.44\linewidth, height=3cm]{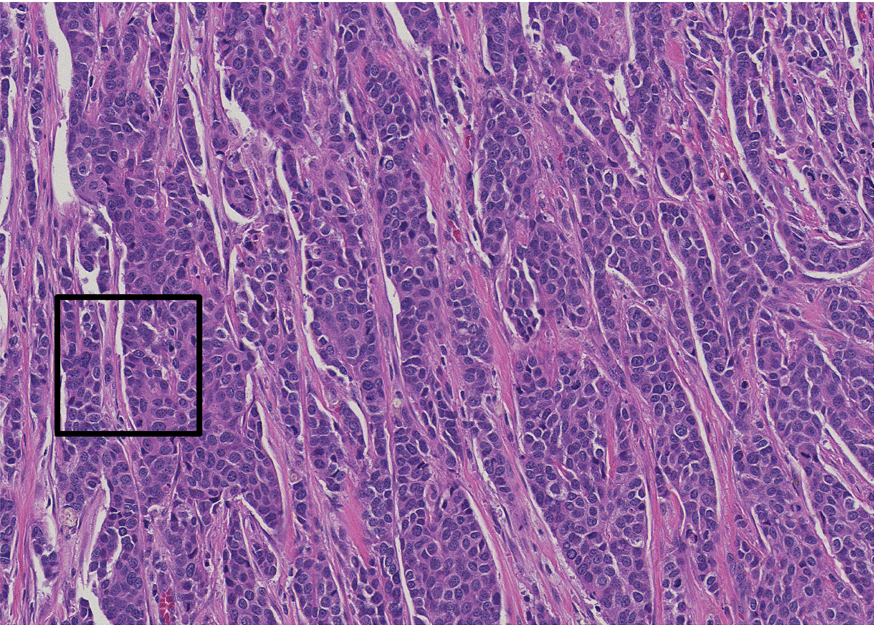}}
  \subfigure[]{
    \label{fig:subfig:SSA-seg02} 
    \includegraphics[width=0.44\linewidth, height=3cm]{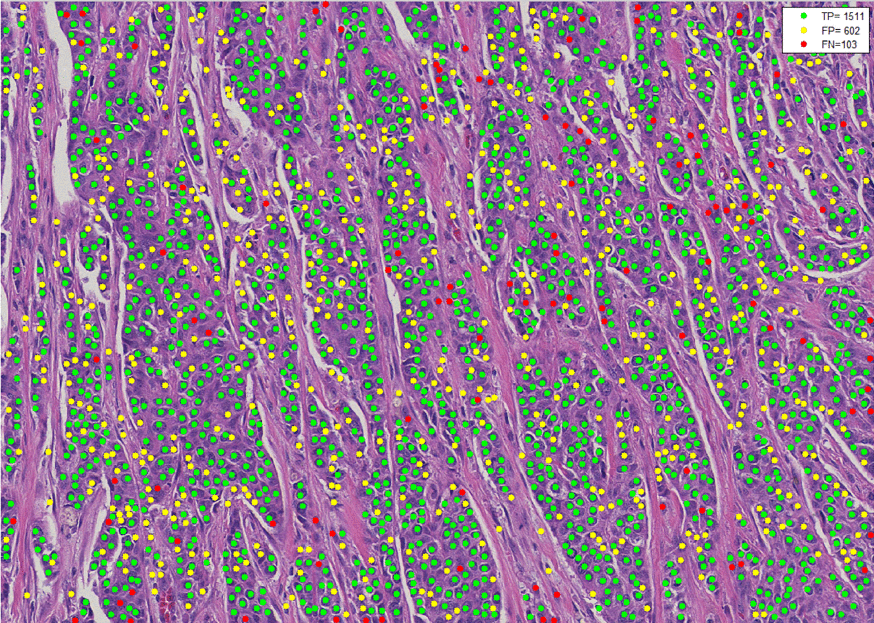}}
     \caption{ An example of nuclei detection results in~\cite{Xu-2016-SSA}. (a) is a whole-slide breast histopathological image, (b) is the nuclei detection results. The green, yellow, and red dots represent the true positive (TP), false positive (FP), and false negative (FN) with respective to groundtruth, respectively. This figure corresponds to Fig.5 in the original paper.}
  \label{fig:SSA-seg} 
\end{figure}

In~\cite{Xu-2016-SSA}, a deep learning strategy named ``Stacked Sparse Auto-Encoder'' (SSAE) is presented to detect the nuclei on high-resolution breast cancer images. In the experiment, SSAE is combined with a softmax classifier (SMC) for nuclei detection. Through the sliding window scheme, randomly selected image patches from histopathological images are fed to the trained SSAE \& SMC model to detect the presence of nuclei. If a nucleus is found, a green dot is placed in the center of the corresponding image patch. The qualitative detection results of SSAE \& SMC for a WSI are shown in \figurename~\ref{fig:SSA-seg}. As a result, an F-measure of $84.49 \%$ and a precision of $ 88.84\%$ are obtained.

In~\cite{Litjens-2016-DLA}, a DCNN model is proposed to detect breast cancer metastasis in sentinel lymph nodes. In the experiment, 100 examples are used for training, 50 for validation, and 75 for testing. Finally, a sensitivity of $99.9\%$ is achieved.

In~\cite{CruzRoa-2018-HASW}, a novel accurate and high-throughput method (HASHI) for automatic invasive breast cancer detection in WSIs is presented. The test is conducted in three different data queues involving 500 cases. Finally, the comparison results of intensive sampling (6 million samples in 24 hours) and less samples (2000 samples in 1 minute) are obtained, an average DC reaches $76\%$ on the independent test dataset.

In~\cite{Monjoy-2018-EDLM}, handcrafted features are combined with high-level features based on deep learning, which are directly fed back to the first fully connected layer for mitotic detection. In this work, three datasets are used, including the MITOS-ATYPIA-14 dataset, ICPR-2012 dataset, and AMIDA-13 dataset. Finally, a precision of $92\%$, a recall of $88\%$ and an F-score of $90\%$ are obtained.

In~\cite{Zainudin-2019-DLCA}, a DCNN architecture is introduced to detect mitosis from histopathological images of breast cancer cells. The data set of MITOS atypia is tested. The result shows that the deeper CNN layer has better the performance of breast cancer image detection. In the 17 layer CNN architecture, $84.49\%$ accuracy, $80.55\%$ TPR, $11.66\%$ FNR and $15.50\%$ loss are achieved on average.

In~\cite{beevi-2019-AMDB}, CNN based deep transfer learning is proposed to achieve automatic mitosis detection in breast histopathology images. A pre-trained VGGNet is transformed by coupling random forest classifier with the initial fully connected layers to extract discriminant features from nuclei patches and to precisely prognosticate the class label of cell nuclei. Finally, average F-score of $88.6\%$ and $89.66\%$ are obtained on MITOS-ATYPIA-14 dataset and RCC dataset (a clinical dataset from Regional Cancer Centre, Thiruvananthapuram, India), respectively.

\subsection{Summary}
\label{ss:deep:summary}
From the survey above, we can find that deep ANN has been increasingly used in the field of BHIA since 2012. Among them, the method based on CNN is dominant. The main reasons for this development trend are as follows: (a) The emergence of high-performance GPU computing makes it possible to train networks with more layers. (b) More and more institutions have released datasets of breast histopathological images, to a certain extent alleviating the lack of labeled public datasets. The large increase in training data reduces the risk of over-fitting. (c) Compared with traditional image classification methods, deep learning can automatically learn features from data, avoiding the complexity and limitations of artificial design and feature extraction in traditional algorithms. (d) CNN has been widely applied in natural language processing, object recognition, image classification, and recognition, laying a foundation for the application of CNN to histopathological images of breast cancer. The work of different teams in the analysis of breast histopathological images using deep neural networks is summarized in \tablename~\ref{table3}. Further details of the method analysis are discussed in Sec.~\ref{ss:methods_deep} and Sec.~\ref{ss:methods_outstanding}.

\onecolumn
\begin{landscape}\scriptsize
\renewcommand\arraystretch{1.85}
\setlength{\tabcolsep}{1.3 pt}
\centering
\begin{longtable}{|c|c|c|c|c|l|l|l|l|}
\caption{Summary of reviewed works of deep neural network methods for BHIA tasks. Sensitivity (Sn), Specificity (Sp), Area Under the Curve (AUC), \\ Precision (P), Recall (R), Accurcy (Acc), Average Accurcy (A-Acc), Dice Coefficient (DC), Average Dice Coefficient (A-DC), Classification (C), Segmentation (S), Detection (D). The third column ``Detail'', shows the number of classes, where ``mul'' stands for multi-class.} \\
\endfirsthead
\caption[l]{Continue: Summary of reviewed works of deep neural network methods for BHIA tasks.}\\
\hline
\endhead
 \hline
\endfoot
\hline
Task                           & Aim                   & Detail                     & Year                  & Reference                                              & Team                                    & ANN type                                        & \multicolumn{1}{l|}{Evaluation}                                                                                           \\ \hline
\multirow{43}{*}{BreaKHis}     & \multirow{2}{*}{C}    & \multirow{2}{*}{2, mul}    & \multirow{2}{*}{2016} & \multirow{2}{*}{\cite{bayramoglu-2016-DLMI}}              & \multirow{2}{*}{N. Bayramoglu, et al.}  & \multirow{2}{*}{CNN}                            & \multicolumn{1}{l|}{The single-task CNN model: A-Acc = $83.25\%$,}                                                        \\
                               &                       &                            &                       &                                                        &                                         &                                                 & \multicolumn{1}{l|}{The multi-task CNN model: A-Acc = $82.13\%$}                                                          \\ \cline{2-8} 
                               & C                     & 2                          & 2017                  & \cite{Spanhol-2016-BCH,Spanhol-2017-DFF,Spanhol-2018-ABC} & F. Spanhol, et al.                      & LeNet and AlexNet                               & \multicolumn{1}{l|}{Overall Acc = $85\%$}                                                                                 \\ \cline{2-8} 
                               & C                     & 2                          & 2017                  & \cite{Song-2017-AFV}                                      & Y. Song, et al.                         & Transfer Learning based VGG-VD                  & \multicolumn{1}{l|}{Acc = $87\%$}                                                                                         \\ \cline{2-8} 
                               & C                     & 2                          & 2017                  & \cite{Zhi-2017-UTL}                                       & W. Zhi, et al.                          & Transfer Learning based VGGNet                  & \multicolumn{1}{l|}{Acc = $90\%$}                                                                                         \\ \cline{2-8} 
                               & C                     & 2                          & 2017                  & \cite{Nejad-2017-COH}                                     & E. Nejad, et al.                        & CNN                                             & \multicolumn{1}{l|}{Acc = $77.5\%$}                                                                                       \\ \cline{2-8} 
                               & C                     & 2                          & 2017                  & \cite{Li-2017-UDL}                                        & Q. Li, et al.                           & CNN                                             & \multicolumn{1}{l|}{Acc = $90\%$}                                                                                         \\ \cline{2-8} 
                               & C                     & 2                          & 2017                  & \cite{das-2017-CHWU}                                      & K. Das, et al.                          & DCNN                                            & \multicolumn{1}{l|}{A-Acc = $94.67\%$, Sn = $96\%$, Sp = $92\%$, F-score = $96.24\%$}                                     \\ \cline{2-8} 
                               & C                     & 2                          & 2017                  & \cite{wei-2017-DLMB}                                      & B. Wei, et al.                          & DCNN                                            & \multicolumn{1}{l|}{Acc = $97\%$}                                                                                         \\ \cline{2-8} 
                               & C                     & 2                          & 2018                  & \cite{Motlagh-2018-BCH}                                   & N. Motlagh, et al.                      & ResNet                                          & \multicolumn{1}{l|}{Acc = $99.6\%$}                                                                                       \\ \cline{2-8} 
                               & \multirow{2}{*}{C}    & \multirow{2}{*}{2}         & \multirow{2}{*}{2018} & \multirow{2}{*}{\cite{Mehra-2018-BCH}}                   & \multirow{2}{*}{R. Mehra, et al.}         & Transfer Learning based                         & \multicolumn{1}{l|}{\multirow{2}{*}{VGG-16 obtains the best Acc = $92.6\%$}}                                              \\
                               &                       &                            &                       &                                                        &                                         & VGG-16, VGG-19, and ResNet-50                   & \multicolumn{1}{l|}{}                                                                                                     \\ \cline{2-8} 
                               & \multirow{2}{*}{C}    & \multirow{2}{*}{mul}       & \multirow{2}{*}{2018} & \multirow{2}{*}{\cite{Nawaz-2018-ACO}}                    & \multirow{2}{*}{M. Nawaz, et al.}       & Inception-V1, Inception-V2                      & \multicolumn{1}{l|}{\multirow{2}{*}{ResNet-V1-50 obtains the highest Acc = $95\%$.}}                                      \\
                               &                       &                            &                       &                                                        &                                         & and ResNet-V1-50                                & \multicolumn{1}{l|}{}                                                                                                     \\ \cline{2-8} 
                               & \multirow{3}{*}{C}    & \multirow{3}{*}{2}         & 2018                  & \cite{Nahid-2018-HBC2}                                    & \multirow{3}{*}{A. Nahid, et al.}       & DCNN                                            & \multicolumn{1}{l|}{The best A-Acc = $87.75\%$}                                                                           \\ \cline{4-5} \cline{7-8} 
                               &                       &                            & 2018                  & \cite{Nahid-2018-HBC}                                     &                                         & CNN and RNN                                     & \multicolumn{1}{l|}{Acc = $91\%$}                                                                                         \\ \cline{4-5} \cline{7-8} 
                               &                       &                            & 2018                  & \cite{Nahid-2018-HBC3}                                    &                                         & DCNN                                            & \multicolumn{1}{l|}{Acc = $92.19\%$}                                                                                      \\ \cline{2-8} 
                               & C                     & 2                          & 2018                  & \cite{Du-2018-BCH}                                        & B. Du, et al.                           & CNN                                             & \multicolumn{1}{l|}{Acc = $90\%$}                                                                                         \\ \cline{2-8} 
                               & C                     & mul                        & 2018                  & \cite{Nawaz-2018-MBC}                                     & M. Nawaz, et al.                        & DenseNet based CNN                              & \multicolumn{1}{l|}{Acc = $95.4\%$}                                                                                       \\ \cline{2-8} 
                               & C                     & 2                          & 2018                  & \cite{Gandomkar-2018-AFF}                                 & Z. Gandomkar, et al.                    & ResNet                                          & \multicolumn{1}{l|}{Acc = $98.77\%$}                                                                                      \\ \cline{2-8} 
                               & \multirow{2}{*}{C}    & \multirow{2}{*}{2, mul}    & \multirow{2}{*}{2018} & \multirow{2}{*}{\cite{DalalBardou-2018-CBCB}}             & \multirow{2}{*}{D. Bardou, et al.}      & \multirow{2}{*}{CNN}                            & \multicolumn{1}{l|}{Binary classification: Acc = between $96.15\%$ and $98.33\%$,}                                        \\
                               &                       &                            &                       &                                                        &                                         &                                                 & \multicolumn{1}{l|}{Multi-class classification: Acc = between $83.31\%$ and $88.23\%$}                                    \\ \cline{2-8} 
                               & \multirow{2}{*}{C}    & \multirow{2}{*}{2}         & \multirow{2}{*}{2018} & \multirow{2}{*}{\cite{KausikDas-2018-MLDC}}               & \multirow{2}{*}{K. Das, et al.}         & \multirow{2}{*}{CNN}                            & \multicolumn{1}{l|}{$40\times $: Acc = $89.52\%$, $100\times $: Acc = $89.06\%$,}                                         \\
                               &                       &                            &                       &                                                        &                                         &                                                 & \multicolumn{1}{l|}{$200\times $: Acc = $88.84\%$, $400\times $: Acc = $87.67\%$}                                         \\ \cline{2-8} 
                               & C                     & 2                          & 2018                  & \cite{Shallu-2018-AMIC}                                   & Shallu, et al.                          & DCNN                                            & \multicolumn{1}{l|}{A-Acc = $85.3\%$}                                                                                     \\ \cline{2-8} 
                               & \multirow{4}{*}{C}    & \multirow{4}{*}{2, mul}    & \multirow{4}{*}{2018} & \multirow{4}{*}{\cite{cascianelli-2018-DRSC}}             & \multirow{4}{*}{S. Cascianelli, et al.} & \multirow{4}{*}{CNN}                            & \multicolumn{1}{l|}{$40\times $: Acc = $87.0\%$, $100\times $: Acc = $85.2\%$,}                                           \\
                               &                       &                            &                       &                                                        &                                         &                                                 & \multicolumn{1}{l|}{$200\times $: Acc = $85.0\%$, $400\times $: Acc = $81.3\%$}                                           \\
                               &                       &                            &                       &                                                        &                                         &                                                 & \multicolumn{1}{l|}{Epistroma dataset: Acc = $94.7\%$}                                                                    \\
                               &                       &                            &                       &                                                        &                                         &                                                 & \multicolumn{1}{l|}{Multi-class Kather's dataset: Acc = $84.0\%$}                                                         \\ \cline{2-8} 
                               & C                     & 2                          & 2019                  & \cite{xu-2019-LICA}                                       & B. Xu, et al.                           & SA-Net                                          & \multicolumn{1}{l|}{A-Acc = $96\%$}                                                                                       \\ \cline{2-8} 
                               & \multirow{3}{*}{C}    & \multirow{3}{*}{2}         & \multirow{3}{*}{2019} & \multirow{3}{*}{\cite{bhuiyan-2019-TLSC}}                 & \multirow{3}{*}{M. Bhuiyan, et al.}     & Transfer Learning based                         & \multicolumn{1}{l|}{\multirow{3}{*}{Acc= $96.24\%$, R = $100\%$, P = $96.55\%$}}                                          \\
                               &                       &                            &                       &                                                        &                                         & ResNet-50, Inception-V2,                        & \multicolumn{1}{l|}{}                                                                                                     \\
                               &                       &                            &                       &                                                        &                                         & Inception ResNet-V2 and Xception                & \multicolumn{1}{l|}{}                                                                                                     \\ \cline{2-8} 
                               & \multirow{2}{*}{C}    & \multirow{2}{*}{2, mul}    & \multirow{2}{*}{2019} & \multirow{2}{*}{\cite{xie-2019-DLBA}}                     & \multirow{2}{*}{J. Xie, et al.}         & \multirow{2}{*}{DCNN}                           & \multicolumn{1}{l|}{Binary classification: Acc = $97.90\%$,}                                                              \\
                               &                       &                            &                       &                                                        &                                         &                                                 & \multicolumn{1}{l|}{Multi-class classification: Acc = $92.07\%$}                                                          \\ \cline{2-8} 
                               & \multirow{2}{*}{C}    & \multirow{2}{*}{2, mul}    & \multirow{2}{*}{2019} & \multirow{2}{*}{\cite{YunJiang-2019-BCH}}                 & \multirow{2}{*}{Y. Jiang, et al.}       & \multirow{2}{*}{BHCNet}                         & \multicolumn{1}{l|}{Binary classification: Acc = between $98.87\%$ and $99.34\%$}                                         \\
                               &                       &                            &                       &                                                        &                                         &                                                 & \multicolumn{1}{l|}{Multi-class classification: Acc = between $90.66\%$ and $93.81\%$}                                    \\ \cline{2-8} 
                               & \multirow{2}{*}{C}    & \multirow{2}{*}{2}         & \multirow{2}{*}{2019} & \multirow{2}{*}{\cite{thuy-2019-FDLT}}                    & \multirow{2}{*}{M. Thuy, et al.}        & Transfer Learning based                         & \multicolumn{1}{l|}{\multirow{2}{*}{Acc = $98.1\%$}}                                                                      \\
                               &                       &                            &                       &                                                        &                                         & VGG-16 and VGG-19                               & \multicolumn{1}{l|}{}                                                                                                     \\ \cline{2-8} 
                               & C                     & 2                          & 2019                  & \cite{Matos-2019-DTLB}                                    & J. Matos, et al.                        & Transfer Learning based Inception-v3            & \multicolumn{1}{l|}{$100\times$: Acc = $91\%$, $200\times$: Acc = $89\%$}                                                 \\ \cline{2-8} 
                               & \multirow{2}{*}{C}    & \multirow{2}{*}{2}         & \multirow{2}{*}{2020} & \multirow{2}{*}{\cite{saxenapre-2020-PCNN}}               & \multirow{2}{*}{S. Saxena, et al.}      & \multirow{2}{*}{CNN}                            & \multicolumn{1}{l|}{$40\times$: Acc = $89.46\%$, $100\times$: Acc = $92.61\%$,}                                           \\
                               &                       &                            &                       &                                                        &                                         &                                                 & \multicolumn{1}{l|}{$200\times$: Acc = $93.92\%$, $400\times$: Acc = $89.78\%$}                                           \\ \cline{2-8} 
                               & C                     & 2                          & 2020                  & \cite{gour-2020-RLBC}                                     & M. Gour, et al.                         & ResHist                                         & \multicolumn{1}{l|}{Acc = $92.52\%$, F1-score = $93.45\%$}                                                                \\ \hline
\multirow{4}{*}{Camelyon 2016} & D                     &                            & 2017                  & \cite{Liu-2017-DCM}                                       & Y. Liu, et al.                          & CNN                                             & \multicolumn{1}{l|}{AUC = $97\%$}                                                                                         \\ \cline{2-8} 
                               & D                     &                            & 2016                  & \cite{Wang-2016-DLF}                                      & D. Wang, et al.                         & GoogLeNet                                       & \multicolumn{1}{l|}{AUC = $92.5\%$}                                                                                       \\ \cline{2-8} 
                               & D                     &                            & 2018                  & \cite{BenTaieb-2018-PCW}                                  & A. BenTaieb, et al.                     & Recurrent Visual Attention Model                & \multicolumn{1}{l|}{AUC = $96\%$}                                                                                         \\ \cline{2-8} 
                               & D                     &                            & 2018                  & \cite{lin-2018-SAFD}                                      & H. Lin, et al.                          & ScanNet                                         & \multicolumn{1}{l|}{FROC score = 0.8533, AUC = $98.75\%$}                                                                 \\ \hline
Camelyon 2017                  & C                     & 4                          & 2017                  & \cite{Chervony-2017-FCO}                                  & L. Chervony, et al.                     & CNNs                                            & \multicolumn{1}{l|}{Acc = $92\%$}                                                                                         \\ \hline
\multirow{23}{*}{BACH}         & C                     & 4                          & 2018                  & \cite{Golatkar-2018-COB}                                  & A. Golatkar, et al.                     & Inception-v3                                    & \multicolumn{1}{l|}{Acc = $85\%$}                                                                                         \\ \cline{2-8} 
                               & C                     & 4                          & 2018                  & \cite{Nazeri-2018-TCN}                                    & K. Nazeri, et al.                       & CNN                                             & \multicolumn{1}{l|}{Acc = $94\%$}                                                                                         \\ \cline{2-8} 
                               & C                     & 4                          & 2018                  & \cite{Kiambe-2018-BHI}                                    & K. Kiambe, et al.                       & AlexNet                                         & \multicolumn{1}{l|}{Acc = $99.84\%$}                                                                                      \\ \cline{2-8} 
                               & C                     & 4                          & 2018                  & \cite{Ranjan-2018-HAF}                                    & N. Ranjan, et al.                       & AlexNet                                         & \multicolumn{1}{l|}{Acc = $95\%$}                                                                                         \\ \cline{2-8} 
                               & \multirow{2}{*}{C}    & \multirow{2}{*}{4}         & \multirow{2}{*}{2018} & \multirow{2}{*}{\cite{Vesal-2018-CBCH}}                   & \multirow{2}{*}{S. Vesal, et al.}       & Transfer Learning based                         & \multicolumn{1}{l|}{Inception-V3: A-Acc = $97.08\%$,}                                                                     \\
                               &                       &                            &                       &                                                        &                                         & Inception-V3 and ResNet-50                      & \multicolumn{1}{l|}{ResNet-50: A-Acc= $96.66\%$}                                                                          \\ \cline{2-8} 
                               & \multirow{2}{*}{C}    & \multirow{2}{*}{4}         & \multirow{2}{*}{2018} & \multirow{2}{*}{\cite{Ferreira-2018-CBCH}}                & \multirow{2}{*}{C. Ferreira, et al.}    & \multirow{2}{*}{Inception ResNet-V2}            & \multicolumn{1}{l|}{Test: Acc = $90\%$, loss = 0.59,}                                                                     \\
                               &                       &                            &                       &                                                        &                                         &                                                 & \multicolumn{1}{l|}{Validation: Acc = $93\%$, loss = 0.23}                                                                \\ \cline{2-8} 
                               & \multirow{2}{*}{C\&S} & \multirow{2}{*}{}          & \multirow{2}{*}{2018} & \multirow{2}{*}{\cite{Vu-2018-MMBH}}                      & \multirow{2}{*}{Q. Vu, et al.}          & \multirow{2}{*}{Encoder and decoder}            & \multicolumn{1}{l|}{Patch classification: Acc = $71\% $(train), Acc = $65\%$(test),}                                      \\
                               &                       &                            &                       &                                                        &                                         &                                                 & \multicolumn{1}{l|}{Segmentation: overall score = 0.7343 (train), overall score = 0.4945 (test)}                          \\ \cline{2-8} 
                               & C\&S                  &                            & 2018                  & \cite{MK-2018-ABCH}                                       & M. Kohl, et al.                         & Densenet-161                                    & \multicolumn{1}{l|}{Acc = $96.24\%$}                                                                                      \\ \cline{2-8} 
                               & C                     & 4                          & 2018                  & \cite{Wang-2018-BCMI}                                     & Y. Wang, et al.                         & CNN                                             & \multicolumn{1}{l|}{Acc = $83\%$}                                                                                         \\ \cline{2-8} 
                               & C                     & 4                          & 2018                  & \cite{Awan-2018-CLUT}                                     & R. Awan, et al.                         & ResNet                                          & \multicolumn{1}{l|}{A-Acc = $83\%$}                                                                                       \\ \cline{2-8} 
                               & \multirow{2}{*}{C}    & \multirow{2}{*}{4}         & \multirow{2}{*}{2018} & \multirow{2}{*}{\cite{Cao-2018-IPTL}}                     & \multirow{2}{*}{H. Cao, et al.}         & Transfer Learning based ResNet-18,              & \multicolumn{1}{l|}{\multirow{2}{*}{A-Acc = $87.1\%$}}                                                                    \\
                               &                       &                            &                       &                                                        &                                         & ResNeXt, NASNet-A,ResNet-152 and VGG-16         & \multicolumn{1}{l|}{}                                                                                                     \\ \cline{2-8} 
                               & C                     & 4                          & 2018                  & \cite{Vang-2018-DLFM}                                     & Y. Vang, et al.                         & Inception-V3                                    & \multicolumn{1}{l|}{A-Acc = $87.5\%$}                                                                                     \\ \cline{2-8} 
                               & C                     & 4                          & 2019                  & \cite{yan-2019-BCHI}                                      & R. Yan, et al.                          & Transfer Learning based Inception-V3            & \multicolumn{1}{l|}{A-Acc = $91.3\%$}                                                                                     \\ \cline{2-8} 
                               & \multirow{2}{*}{C}    & \multirow{2}{*}{2, 4}      & \multirow{2}{*}{2019} & \multirow{2}{*}{\cite{K.Roy-2019-PBSC}}                   & \multirow{2}{*}{K. Roy, et al.}         & \multirow{2}{*}{CNN}                            & \multicolumn{1}{l|}{2-class classification: Acc = $92.5\%$,}                                                              \\
                               &                       &                            &                       &                                                        &                                         &                                                 & \multicolumn{1}{l|}{4-class classification: Acc = $90\%$}                                                                 \\ \cline{2-8} 
                               & C                     &                            & 2019                  & \cite{kassani-2019-BCDT}                                  & S. Kassani, et al.                      & Transfer Learning based Xception                & \multicolumn{1}{l|}{A-Acc = $92.50\%$}                                                                                    \\ \cline{2-8} 
                               & \multirow{3}{*}{C}    & \multirow{3}{*}{2, 4, mul} & \multirow{3}{*}{2019} & \multirow{3}{*}{\cite{kausar-2019-MRBH}}                  & \multirow{3}{*}{T. Kausar, et al.}      & \multirow{3}{*}{DCNN}                           & \multicolumn{1}{l|}{BACH: 2-class classification: Acc = $98.2\%$,}                                                        \\
                               &                       &                            &                       &                                                        &                                         &                                                 & BACH: 4-class classification: Acc = \$98.2\%,                                                                             \\
                               &                       &                            &                       &                                                        &                                         &                                                 & BreaKHis: multi-class classification: Acc = $96.85\%$                                                                     \\ \hline
\multirow{8}{*}{ICPR 2012}     & D                     &                            & 2013                  & \cite{roux-2013-MDBC}                                     & L. Roux, et al.                         & CNN                                             & \multicolumn{1}{l|}{R = $70\%$ , Acc = $89\%$, F-measure = $78\%$}                                                        \\ \cline{2-8} 
                               & D\&C                  &                            & 2013                  & \cite{Ciresan-2013-MDI}                                   & D. Ciresan, et al.                      & DCNN                                            & \multicolumn{1}{l|}{F1-score = $78.2\%$}                                                                                  \\ \cline{2-8} 
                               & S                     &                            & 2014                  & \cite{Veta-2014-BCH2}                                     & M. Veta, et al.                         & DCNN                                            & \multicolumn{1}{l|}{F1-score = $61.1\%$}                                                                                  \\ \cline{2-8} 
                               & C                     & \multirow{2}{*}{2}         & 2012                  & \cite{Malona-2012-MFR}                                    & \multirow{2}{*}{C. Malon, et al.}       & \multirow{2}{*}{CNN}                            & \multicolumn{1}{l|}{Color scanners: F1-score = $65.9\%$,}                                                                 \\ \cline{2-2} \cline{4-5}
                               & C                     &                            & 2013                  & \cite{Malon-2013-COM}                                     &                                         &                                                 & \multicolumn{1}{l|}{Multi-spectral scanners: F1-score = $58.9\%$}                                                         \\ \cline{2-8} 
                               & D                     &                            & 2014                  & \cite{Wang-2014-CEO}                                      & H. Wang, et al.                         & DCNN                                            & \multicolumn{1}{l|}{F1-score = $73.5\%$}                                                                                  \\ \cline{2-8} 
                               & \multirow{2}{*}{D}    & \multirow{2}{*}{}          & \multirow{2}{*}{2016} & \multirow{2}{*}{\cite{chen-2016-MDBC}}                    & \multirow{2}{*}{H. Chen, et al.}        & \multirow{2}{*}{CasCNN}                         & \multicolumn{1}{l|}{ICPR12: P = $80.4\%$, R = $72.2\%$, F1-score = $78.8\%$,}                                             \\
                               &                       &                            &                       &                                                        &                                         &                                                 & \multicolumn{1}{l|}{ICPR14: P = $46\%$, R = $50.7\%$, F1-score = $48.2\%$}                                                \\ \hline
\multirow{4}{*}{TCUG16}        & S\&D                  &                            & 2019                  & \cite{Wahab-2019-TLBD}                                    & N. Wahab, et al.                        & Transfer Learning based DCNN                    & \multicolumn{1}{l|}{F-measure = $71.3\%$}                                                                                 \\ \cline{2-8} 
                               & D                     &                            & 2019                  & \cite{ChaoLi-2019-WSMD}                                   & C. Li, et al.                           & Deep cascade CNN                                & \multicolumn{1}{l|}{F-score = $66.9\%$}                                                                                   \\ \cline{2-8} 
                               & \multirow{2}{*}{D}    & \multirow{2}{*}{}          & \multirow{2}{*}{2019} & \multirow{2}{*}{\cite{veta-2019-PBTP}}                    & \multirow{2}{*}{M. Veta, et al.}        & \multirow{2}{*}{DCNNs}                          & \multicolumn{1}{l|}{Task 1: $k$ = 0.567, $95\%$ CI {[}0.464, 0.671{]} between the predicted scores and the ground truth.} \\
                               &                       &                            &                       &                                                        &                                         &                                                 & \multicolumn{1}{l|}{Task 2: $r$ = 0.617, $95\%$ CI {[}0.581 0.651{]} with the ground truth.}                              \\ \hline
                               & C                     & 4                          & 2017                  & \cite{Araujo-2017-COB}                                    & T. Araujo, et al.                       & DCNN                                            & \multicolumn{1}{l|}{Acc = $77.8\%$}                                                                                       \\ \cline{2-8} 
                               & \multirow{2}{*}{C}    & \multirow{2}{*}{4}         & \multirow{2}{*}{2018} & \multirow{2}{*}{\cite{Mahbod-2018-BCH}}                   & \multirow{2}{*}{A. Mahbod, et al.}      & \multirow{2}{*}{ResNet-50 and ResNet-101}       & \multicolumn{1}{l|}{Acc = $97.22\%$,}                                                                                     \\
Bioimaging                     &                       &                            &                       &                                                        &                                         &                                                 & \multicolumn{1}{l|}{BACH dataset: Acc = $88.5\%$}                                                                         \\ \cline{2-8} 
2015                           & \multirow{2}{*}{C}    & \multirow{2}{*}{2, 4}      & \multirow{2}{*}{2018} & \multirow{2}{*}{\cite{Rakhlin-2018-DCN}}                  & \multirow{2}{*}{A. Rakhlin, et al.}     & Transfer Learning based                         & \multicolumn{1}{l|}{2-class classification task: Acc = $93.8\%$, AUC = $97.3\%$, Sn = $96.5\%$, Sp = $88\%$,}             \\
Breast                         &                       &                            &                       &                                                        &                                         & ResNet-50, Inception-V3 and VGG-16              & \multicolumn{1}{l|}{4-class classification task: Acc = $87.2\%$}                                                          \\ \cline{2-8} 
Histology                      & C                     & 4                          & 2019                  & \cite{YuqianLi-2019-COBC}                                 & Y. Li, et al.                           & CNN, ResNet-50                                  & \multicolumn{1}{l|}{Acc = $95\%$}                                                                                         \\ \cline{2-8} 
Classification                 & \multirow{2}{*}{C}    & \multirow{2}{*}{2, mul}    & \multirow{2}{*}{2019} & \multirow{2}{*}{\cite{Md-2019-BCCH}}                      & \multirow{2}{*}{M. Alom, et al.}        & \multirow{2}{*}{IRRCNN}                         & \multicolumn{1}{l|}{Binary classification: Acc = $99.05\%$,}                                                              \\
Challenge                      &                       &                            &                       &                                                        &                                         &                                                 & \multicolumn{1}{l|}{Multi-class classification: Acc = $98.59\%$}                                                          \\ \cline{2-8} 
                               & \multirow{2}{*}{C}    & \multirow{2}{*}{4}         & \multirow{2}{*}{2019} & \multirow{2}{*}{\cite{Hafiz-2019-COBC}}                   & \multirow{2}{*}{H. Ahmad, et al.}       & Transfer Learning based                         & \multicolumn{1}{l|}{Pach level: the best Acc = $83.6\%$,}                                                                 \\
                               &                       &                            &                       &                                                        &                                         & AlexNet, GoogleNet, and ResNet                  & \multicolumn{1}{l|}{Image level: the best Acc = $85\%$}                                                                   \\ \hline
\multirow{2}{*}{IDC}           & D                     &                            & 2014                  & \cite{cruz-2014-ADID}                                     & A. Cruz-Roa, et al.                     & CNN                                             & \multicolumn{1}{l|}{F-measure = $71.80\%$, Acc = $84.23\%$}                                                               \\ \cline{2-8} 
                               & D                     &                            & 2019                  & \cite{FP-2019-MBNI}                                       & F. Romero, et al.                       & CNN                                             & \multicolumn{1}{l|}{F1-score = $90\%$, Acc = $89\%$}                                                                      \\ \hline
\multirow{7}{*}{Others-Class}  & C                     &                            & 2014                  & \cite{Wu-2014-HIC}                                        & J. Wu, et al.                           & PCANet                                          & \multicolumn{1}{l|}{Acc = $79\%$}                                                                                         \\ \cline{2-8} 
                               & C                     & 2                          & 2017                  & \cite{bejnordi-2017-DLAT}                                 & B. Bejnordi, et al.                     & CNN                                             & \multicolumn{1}{l|}{ROC = 0.921}                                                                                          \\ \cline{2-8} 
                               & C                     & 4                          & 2018                  & \cite{Wang-2018-COB}                                      & Z. Wang, et al.                         & VGG-16 and VGG-19                               & \multicolumn{1}{l|}{A-Acc = $92\%$}                                                                                       \\ \cline{2-8} 
                               & C                     & mul                        & 2018                  & \cite{Gecer-2018-DAC}                                     & B. Gecer, et al.                        & CNN                                             & \multicolumn{1}{l|}{Acc = $55\%$}                                                                                         \\ \cline{2-8} 
                               & C                     & mul                        & 2018                  & \cite{HD-2018-IAWD}                                       & H. Couture, et al.                      & CNN                                             & \multicolumn{1}{l|}{A-Acc = $82.4\%$}                                                                                     \\ \cline{2-8} 
                               & \multirow{2}{*}{C}    & \multirow{2}{*}{2}         & \multirow{2}{*}{2019} & \multirow{2}{*}{\cite{khan-2019-ANDL}}                    & \multirow{2}{*}{S. Khan, et al.}        & Transfer Learning based                         & \multicolumn{1}{l|}{\multirow{2}{*}{A-Acc = $97.525\%$}}                                                                  \\
                               &                       &                            &                       &                                                        &                                         & GoogLeNet, VGGNet, and ResNet                   & \multicolumn{1}{l|}{}                                                                                                     \\ \hline
\multirow{11}{*}{Others-Seg}   & S                     &                            & 2010                  & \cite{Pang-2010-CNS}                                      & B. Pang, et al.                         & CNN                                             & \multicolumn{1}{l|}{Acc = $95\%$}                                                                                         \\ \cline{2-8} 
                               & \multirow{2}{*}{S}    & \multirow{2}{*}{}          & \multirow{2}{*}{2015} & \multirow{2}{*}{\cite{su-2015-RSHB}}                      & \multirow{2}{*}{H. Su, et al.}          & \multirow{2}{*}{fCNN}                           & \multicolumn{1}{l|}{One $1000\times 1000$ image can be segmented in 2.3 seconds,}                                         \\
                               &                       &                            &                       &                                                        &                                         &                                                 & \multicolumn{1}{l|}{Mean: P = $91\%$, R = $82\%$, F1-score = $85\%$}                                                      \\ \cline{2-8} 
                               & \multirow{2}{*}{S}    & \multirow{2}{*}{}          & \multirow{2}{*}{2016} & \multirow{2}{*}{\cite{xu-2016-ADCN}}                      & \multirow{2}{*}{J. xu, et al.}          & \multirow{2}{*}{DCNN}                           & \multicolumn{1}{l|}{NKI dataset: F-score = $85.21\%$, Acc = $84.34\%$,}                                                   \\
                               &                       &                            &                       &                                                        &                                         &                                                 & \multicolumn{1}{l|}{VGH dataset: F-score = $89.10\%$, Acc = $88.34\%$}                                                    \\ \cline{2-8} 
                               & S                     &                            & 2016                  & \cite{xing-2016-AALB}                                     & F. Xing, et al.                         & DCNN                                            & \multicolumn{1}{l|}{Acc = $92\%$}                                                                                         \\ \cline{2-8} 
                               & S                     &                            & 2017                  & \cite{pan-2017-ASNP}                                      & X. Pan, et al.                          & DCN                                             & \multicolumn{1}{l|}{F1-measure = $83.93\%$, Acc = $92.45\%$}                                                              \\ \cline{2-8} 
                               & S                     &                            & 2017                  & \cite{naylor-2017-NSHI}                                   & P. Naylor, et al.                       & Deep Learning                                   & \multicolumn{1}{l|}{Acc = $95.4\%$, R = $77.3\%$, P = $86.4\%$, F1-score = $80.5\%$}                                      \\ \cline{2-8} 
                               & S                     &                            & 2018                  & \cite{CuiYuxin-2018-ADLA}                                 & Y. Cui, et al.                          & CNN                                             & \multicolumn{1}{l|}{One $1000\times 1000$ image can be segmented in less than 5 seconds}                                  \\ \cline{2-8} 
                               & S                     &                            & 2019                  & \cite{Sonia-2019-DACS}                                    & S. Mejbri, et al.                       & Deep neural networks                            & \multicolumn{1}{l|}{U-Net: DC = 0.86, SegNet: DC = 0.87, FCN: DC = 0.86, DeepLab: DC = 0.86}                              \\ \cline{2-8} 
                               & S                     &                            & 2020                  & \cite{priego-2020-ASWS}                                   & B. Priego-Torres, et al.                & DCNN                                            & \multicolumn{1}{l|}{Acc = $95.62\%$}                                                                                      \\ \hline
\multirow{7}{*}{Others-Det}    & D                     &                            & 2016                  & \cite{Xu-2016-SSA}                                        & J. Xu, et al.                           & Deep learning                                   & \multicolumn{1}{l|}{F-measure = $84.49\%$, P = $88.84\%$}                                                                 \\ \cline{2-8} 
                               & D                     &                            & 2016                  & \cite{Litjens-2016-DLA}                                   & G. Litjens, et al.                      & DCNN                                            & \multicolumn{1}{l|}{Sn = $99.9\%$}                                                                                        \\ \cline{2-8} 
                               & D                     &                            & 2018                  & \cite{CruzRoa-2018-HASW}                                  & A. Cruz-Roa, et al.                     & CNN                                             & \multicolumn{1}{l|}{A-ADC = $76\%$}                                                                                       \\ \cline{2-8} 
                               & D                     &                            & 2018                  & \cite{Monjoy-2018-EDLM}                                   & M. Saha, et al.                         & Deep learning                                   & \multicolumn{1}{l|}{P = $92\%$, R = $88\%$, F-score = $90\%$}                                                             \\ \cline{2-8} 
                               & D                     &                            & 2019                  & \cite{Zainudin-2019-DLCA}                                 & Z. Zainudin, et al.                     & DCNN                                            & \multicolumn{1}{l|}{A-Acc = $84.49\%$}                                                                                    \\ \cline{2-8} 
                               & \multirow{2}{*}{D}    & \multirow{2}{*}{}          & \multirow{2}{*}{2019} & \multirow{2}{*}{\cite{beevi-2019-AMDB}}                   & \multirow{2}{*}{K. Beevi, et al.}       & \multirow{2}{*}{Transfer learning based VGGNet} & \multicolumn{1}{l|}{MITOS-ATYPIA-14 dataset: F-score = $88.60\%$,}                                                        \\
                               &                       &                            &                       &                                                        &                                         &                                                 & \multicolumn{1}{l|}{RCC dataset: F-score = $89.66\%$}                                                                      
\label{table3}\\
\end{longtable}
\end{landscape}
\twocolumn

\section{Methodology Analysis}
\label{s:method} 

An overview of the deeper analysis of classical ANNs and deep ANNs is compiled in this section. Meanwhile, the outstanding methods in different tasks are analyzed.

\subsection{Analysis of classical ANN methods}
\label{ss:methods_classical}

\begin{figure}[htbp!]
\centering
 \includegraphics[width=1\linewidth]{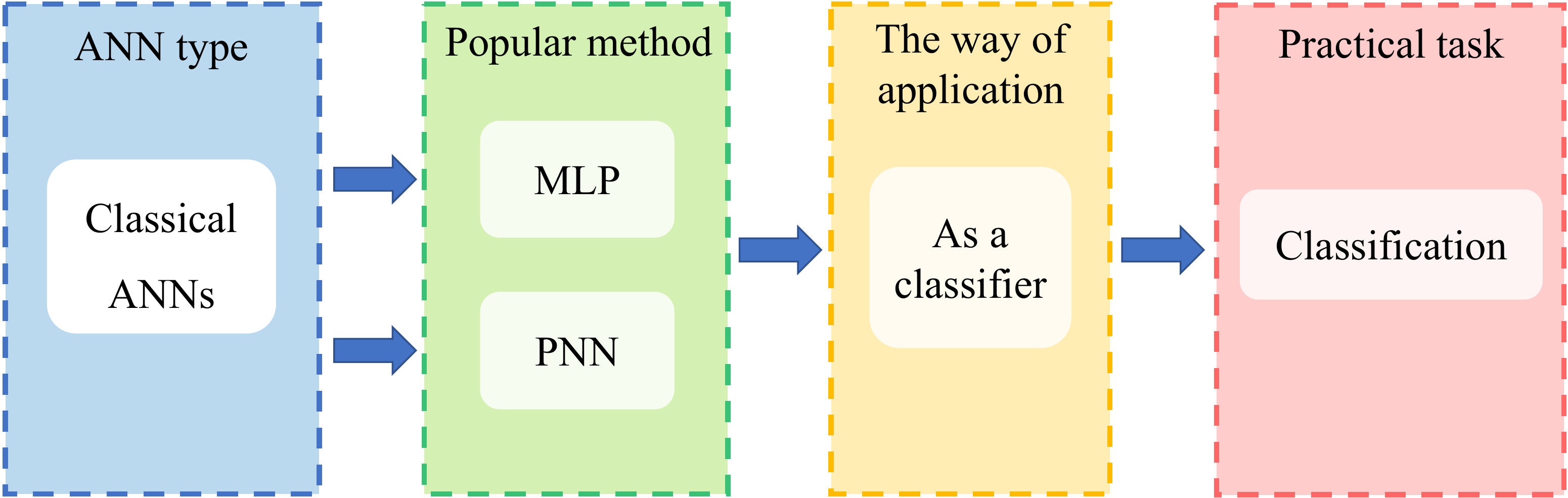}
\caption{The popular methods in classical ANN for BHIA tasks.}
\label{fig:MLP}
\end{figure}

According to the survey on classical ANNs, the Multi-Layer Perceptron (MLP) and Probabilistic Neural Network (PNN) are used more frequently in the analysis of breast histopathological images, as shown in~\figurename\ref{fig:MLP}. Since the datasets employed in each work are different, the evaluation of each method cannot be evaluated longitudinally. Therefore, it is analyzed from the perspective of the neural network itself. MLP is known as a feed-forward neural network, which can solve the problem of linear inseparability and can be trained to accurately generalize when presented with new, unseen data \cite{Gardner-1998-ANNA}. However, the connection mode between its hidden layers is ``fully connected'', which causes too many training parameters. Therefore, It makes it difficult to have too many layers to solve more complex problems. At the same time, the learning speed of MLP is slow and it is easy to fall into local extremes. The papers involved in this article are~\cite{Zhang-2011-BCC}, ~\cite{Zhang-2013-BCD}, ~\cite{Zhang-2013-BCH},~\cite{shukla-2017-CHIB}. PNN is also a kind of feed-forward neural network. Compared with the MLP, the training speed of the PNN is faster and the PNN is usually more accurate than the MLP. The disadvantage is that it is slower than the MLP in classifying new cases and requires additional storage space to store the model. The papers involved in this article are ~\cite{osareh-2010-MLTD}, ~\cite{Loukas-2013-BCC}.

\subsection{Analysis of Deep ANN methods}
\label{ss:methods_deep}

\begin{figure}[htbp!]
\centering
 \includegraphics[width=1\linewidth]{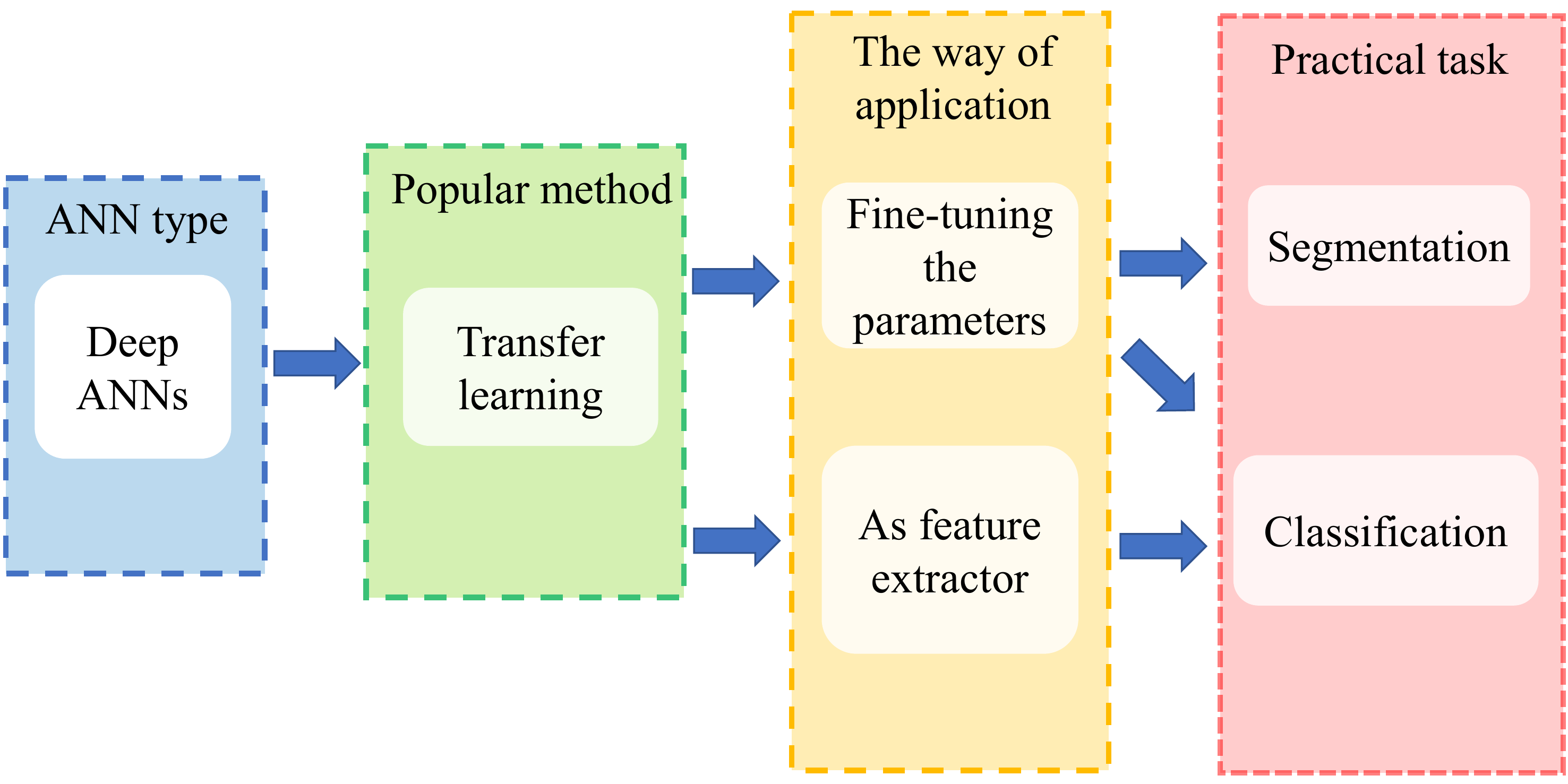}
\caption{The popular methods in deep ANN for BHIA tasks.}
\label{fig:TL}
\end{figure}

In deep ANNs, transfer learning strategies are applied more frequently in the classification of breast histopathological images in recent four years. The papers involved in this article are~\cite{Song-2017-AFV},\cite{Zhi-2017-UTL},\cite{Mehra-2018-BCH},\cite{Nawaz-2018-ACO},\cite{bhuiyan-2019-TLSC},\cite{thuy-2019-FDLT},\cite{Matos-2019-DTLB},\cite{saxenapre-2020-PCNN},\cite{Vesal-2018-CBCH},\cite{Cao-2018-IPTL},\cite{yan-2019-BCHI},\cite{kassani-2019-BCDT},\cite{Wahab-2019-TLBD},\cite{Mahbod-2018-BCH},\cite{Rakhlin-2018-DCN},\cite{Hafiz-2019-COBC},\cite{khan-2019-ANDL},\cite{beevi-2019-AMDB}. Transfer learning is a method used to transfer knowledge acquired from one task to resolve another\cite{Ribani-2019-ASTL}. As shown in in~\figurename\ref{fig:TL}, there are two main approaches for applying transfer learning: (1) Fine-tuning the parameters in the pre-training network according to the required tasks (e.g. \cite{Zhi-2017-UTL},\cite{Nawaz-2018-ACO},\cite{Vesal-2018-CBCH},\cite{Wahab-2019-TLBD},\cite{Mahbod-2018-BCH},\cite{Hafiz-2019-COBC}). (2) Using a pre-trained network as a feature extractor, and then using these features to train a new classifier (e.g.\cite{Song-2017-AFV},\cite{Mehra-2018-BCH},\cite{bhuiyan-2019-TLSC},\cite{thuy-2019-FDLT},\cite{Matos-2019-DTLB},\cite{saxenapre-2020-PCNN},\cite{Cao-2018-IPTL},\cite{yan-2019-BCHI},\cite{kassani-2019-BCDT},\cite{Rakhlin-2018-DCN},\cite{khan-2019-ANDL},\cite{beevi-2019-AMDB}). In the transfer learning, the VGG16~\cite{Simonyan-2014-VDCN}, VGG19, and ResNet50~\cite{He-2016-DRLI} are very popular pre-trained CNN models due to their more in-depth architectures \cite{Mehra-2018-BCH}. The main reasons are as follows: First, due to the inherent complexity and diversity of breast tissue pathological images, it is not easy to label the images, and the cost of labeling data by medical experts is very expensive. Therefore, there are few publicly available large-scale labeled image datasets. However, transfer learning can overcome the problem of small datasets effectively \cite{Hadad-2017-CBLU}. Secondly, in the classification task of breast histopathology images, most of the pre-trained models are from the ImageNet Large Scale Visual Recognition Challenge\cite{Russakovsky-2015-ILSV}. They achieve stable performance on specific tasks and can be safely used for transfer learning in breast cancer classification tasks. Finally, the transfer learning process are helpful to improve accuracy or reduce training time \cite{sarkar-2018-HOTL}, which is an important reason for its popularity.

\subsection{Analysis of the outstanding methods in each reviewed task}
\label{ss:methods_outstanding}

In different review tasks, there are some excellent methods proposed. For example, in the BreakHis dataset task, the best results are obtained in~\cite{YunJiang-2019-BCH}, where a small SE-ResNet model based on the combination of residual module and Squeeze-and-Excitation block is designed, which can effectively reduce model training parameters. Furthermore, a new learning rate scheduler named Gaussian error scheduler is proposed, which can get excellent performance without complicatedly fine-tuning the learning rate. In the Camelyon 2016 dataset task, the best results are obtained in~\cite{lin-2018-SAFD}. In order to detect metastatic breast cancer from WSIs, a fast and dense scanning framework is proposed, referred to as ScanNet. ScanNet is implemented based on the VGG-16 network by changing the last three fully connected layers to fully convolutional layers. In the end, faster performance on tumor localization tasks is achieved and even surpasses human performance on WSI classification tasks. In the ICPR 2012 dataset task, the best results are obtained in~\cite{chen-2016-MDBC}, where a novel deep cascaded convolutional neural network (CasCNN) is designed to detect mitosis. The advantage of the CasCNN is that it can significantly reduce the detection time and achieve satisfactory accuracy. In the Bioimaging 2015 Breast Histology Classification Challenge dataset task, the best results are obtained in \cite{Md-2019-BCCH}, where a method for breast cancer classification with the Inception Recurrent Residual Convolutional Neural Network (IRRCNN) model is proposed. The IRRCNN~\cite{alom-2018-IIRC,alom-2017-IRCN} is one of the improved hybrid DCNN architectures based on inception~\cite{szegedy-2017-IITI}, residual networks~\cite{He-2016-DRLI}, and the RCNN architectures~\cite{liang-2015-RCNN}. Compared with them, the main advantage of this model is that better recognition performance can be achieved using the same number or fewer network parameters.

\subsection{The potential of the methods mentioned in this review in other fields}
\label{ss:methods_potential}

In addition, this review discussed the deep ANNs method not only can be applied in the field of breast histopathological image analysis, but also in the field of other closed microscopic image analysis, such as: Cervical histopathological analysis~\cite{chen-2020-DTHD},\cite{Sornapudi-2018-DLND},\cite{sheikhzadeh-2018-ALMB}, cervical cytopathological analysis~\cite{wu-2018-ACCC},\cite{gautam-2018-CAPS},\cite{Song-2016-ACCS}, stem cell analysis~\cite{padi-2020-CAIB},\cite{kusumoto-2018-ADLB}, microbiological image analysis~\cite{ito-2018-VPDC},\cite{matuszewski-2018-MATS},\cite{kosov-2018-EMCU}, sperm quality analysis~\cite{javadi-2019-ANDL},\cite{riordon-2019-DLTC},\cite{javadi-2019-ANDL}, web-based platform for computer assisted diagnosis~\cite{houston1-999-MDMO},\cite{markiewicz-2016-MIAP}, and rock microstructural analysis~\cite{alqahtani-2018-DLCN},\cite{karimpouli-2019-SDRI}. No matter from the aspects of image pre-processing, feature extraction and selection, segmentation, and classification, or from the aspects of deep ANN model design and proposed framework idea, the methods of deep ANN summarized in this review can bring a new perspective to the research in other fields.

\section{Conclusion and Future Work}
\label{s:conc}
In this review, the methods of breast cancer histopathological image analysis based on the artificial neural network are comprehensively summarized, which are grouped into the classical artificial neural network and deep neural network methods. In addition, when summarizing the deep neural network method, the related work is grouped according to the applied datasets. In each dataset, the related works are arranged in ascending chronological order. From classical review works in Sec.~\ref{s:classical} and subsequent analysis in Sec.~\ref{ss:classical:related}, it is found that the ANNs used in the BHIA field around 2012 are classical neural networks. In the analysis of histopathological images of breast cancer, MLP and PNN are the most applied classical ANNs. However, they are only used as classifiers. In feature extraction, most of the research is used for texture features and morphological features. Among deep learning based methods whose related works and corresponding analysis are discussed in Sec.~\ref{ss:deep:related} and Sec.~\ref{ss:deep:summary}, deep learning technology, especially deep convolutional neural networks, has made excellent achievements in the classification and segmentation of breast histopathological images, which will help patients with early detection, diagnosis, and treatment of breast cancer. According to the survey, transfer learning methods based on CNN are the most frequently used. But from the review works in Sec.~\ref{ss:methods_outstanding}, improved and novel network frameworks tend to perform better in different datasets.

In the future, there is still room for improvement. First, researchers can combine the characteristics of pathological images to develop a new network model to analyze the histopathological images of breast cancer. Secondly, there is still a lack of large-scale, comprehensive, and fully-labeled WSI datasets. Therefore, the establishment of large public datasets is of great value for future research. Thirdly, the classification system of breast cancer is complex, and there are many subtypes under each lesion type. Studying patterns correlated to molecular subtype, treatment response, and prognosis to refine the diagnosis in precision medicine remains a significant challenge~\cite{robertson-2018-DIAB}. Finally, GANs are currently used to generate datasets. However, the advantages of microscopic image analysis have not been explored, which will be a research direction with great potential and value in the future.

\section*{Acknowledgements}
\label{s:ack}
We thank Miss Zixian Li and Mr. Guoxian Li for their important discussion.

\bibliographystyle{IEEEtran}
\bibliography{RefURLLC}

\EOD
\end{document}